\documentclass[fleqn,usenatbib]{mnras}

\usepackage{newtxtext,newtxmath}
\usepackage[T1]{fontenc}

\DeclareRobustCommand{\VAN}[3]{#2}
\let\VANthebibliography\thebibliography
\def\thebibliography{\DeclareRobustCommand{\VAN}[3]{##3}\VANthebibliography}
\usepackage{graphicx}	% Including figure files
\usepackage{amsmath}	% Advanced maths commands

\usepackage{multirow}
\usepackage{xcolor}
\usepackage{caption}
\usepackage{comment}
\usepackage{subcaption}
\usepackage{threeparttable}
\usepackage{rotating}
\usepackage{rotfloat}
\usepackage{float}
\usepackage{moresize}
\usepackage{wrapfig}
\usepackage{multicol}
\usepackage{wrapfig}
\usepackage[normalem]{ulem}
\usepackage[T1]{fontenc}
\usepackage{color}
\usepackage{epstopdf}
\usepackage{longtable}
\usepackage{natbib}
\usepackage[export]{adjustbox}
\usepackage{tabularx}
\usepackage{subcaption}
\usepackage[symbol]{footmisc}
\usepackage{url}
\usepackage{acronym}
\acrodef{LC}[LC]{light curve}
\acrodef{GP}[GP]{Gaussian process}
\acrodef{MCMC}[MCMC]{Markov Chain Monte Carlo}
\acrodef{HJ}[HJ]{Hot Jupiter}
\acrodef{LD}[LD]{limb darkening}
\acrodef{PSD}[PSD]{power spectral density}
\acrodef{CCF}[CCF]{cross-correlation function}
\acrodef{RV}[RV]{radial velocity}
\acrodef{RM}[RM]{Rossiter–McLaughlin}
\acrodef{CLV}[CLV]{center-to-limb variation}
\acrodef{SN}[S/N]{signal-to-noise ratio}
\acrodef{VMR}[VMR]{volume mixing ratio}

\newcommand{\vsini}{$v\sin{i_\star}$}

\newcommand{\vsys}{$v_{\rm sys}$}
\newcommand{\teff}{$T_{\rm eff}$}
\newcommand{\logg}{log\,{\it g$_\star$}}

\newcommand{\feh}{[Fe/H]}

\renewcommand{\arraystretch}{1.3}

\newcommand{\kms}{km\,s$^{-1}$}
\newcommand{\ms}{m~s$^{-1}$}

\newcommand{\Msun}{$M_{\odot}$}
\newcommand{\Rsun}{$R_{\odot}$}

\newcommand{\MJ}{$M_\mathrm{J}$}
\newcommand{\RJ}{$R_\mathrm{J}$}

\newcommand{\rstar}{$R_\star$}

\newcommand{\rplanet}{$R_{\mathrm{p}}$}
\newcommand{\mplanet}{$M_{\mathrm{p}}$}
\newcommand{\Teq}{$T_\mathrm{eq}$}
\newcommand{\harpsn}{HARPS-N}
\newcommand{\logrhk}{$\rm log\,R^{\prime}_\mathrm{HK}$}

\title[WASP-13\,b with GIARPS]{The GAPS programme at TNG\\ LXXXII. The first atmospheric investigation of the gas giant WASP-13\,b at high resolution with GIARPS\thanks{Based on observations made with the Italian Telescopio Nazionale Galileo (TNG) operated on the island of La Palma by the Fundación Galileo Galilei (FGG) of the Istituto Nazionale di Astrofisica (INAF) at the Spanish Observatorio del Roque de los Muchachos of the Instituto de Astrofisica de Canarias.}}

\author[Sicilia, Basilicata et al.]{D. Sicilia$^{1}$\thanks{E-mail: daniela.sicilia@inaf.it},
M. Basilicata$^{2}$,
G. Guilluy$^{3}$,
G. Scandariato$^{1}$,
F. Borsa$^{2}$,
C. Di Maio$^{4}$,
M. Brogi$^{5,3}$,
\newauthor
A. Chiavassa$^{6,3}$,
L. Fossati$^{7}$,
F. Amadori$^{3,5}$,
P. Giacobbe$^{3}$,
A. S. Bonomo$^{3}$,
A. Sozzetti$^{3}$,
L. Mancini$^{8,3}$,
\newauthor
G. Andreuzzi$^{9,10}$,
S. Colombo$^{4}$,
V. Fardella$^{4,11}$,
A. Ghedina$^{9}$,
G. Mantovan$^{12,13}$ 
and G. Piccinini$^{4,11}$\\
\\
Affiliations are listed at the end of the paper}

\date{Accepted XXX. Received YYY; in original form ZZZ}
\pubyear{\the\year{}}
\begin{document}
\label{firstpage}
\pagerange{\pageref{firstpage}--\pageref{lastpage}}
\maketitle  
\begin{abstract}
    Low-density transiting gas giant exoplanets are key benchmarks for atmospheric studies. Despite their large-scale heights, many of them remain poorly characterised. We present the first atmospheric investigation at high-resolution of the inflated (R$_\mathrm{p}$$\sim$1.4\,R$_{\mathrm{J}}$), hot Jupiter (T$\sim$1500\,$\mathrm{K}$) WASP-13\,b aiming at constraining its atmospheric composition and refine its orbital geometry. We analysed four transits observed with GIARPS at the TNG. Updated stellar and planetary parameters were re-derived from the most precise literature values, and a new Rossiter-McLaughlin analysis was performed to measure the projected spin-orbit angle ($\lambda$) and stellar rotational velocity ($v\sin{i_\star}$). Atmospheric signals were searched using high-resolution spectroscopy, both with cross-correlation and single-line analyses, targeting key atomic and molecular species (e.g. \ion{He}{I}, H$\alpha$, \ion{Na}{I}, H$_2$O, CO). We find a well-aligned orbit ($\lambda$\,=\,2.7$\pm$1.1$^\circ$) and $v\sin{i_\star}$\,=\,4.5$\pm$0.1 km\,s$^{-1}$, improving upon previous constraints. No significant atmospheric absorption is detected in either VIS or NIR data. Injection-and-recovery tests exclude an isothermal atmosphere with metallicity [M/H]$\lesssim$10$\times$[M/H]$_\odot$ and simultaneously a grey cloud top pressure P$_{\rm cloud}\gtrsim10$\,mbar, under the assumption of chemical equilibrium. WASP-13\,b appears to belong to the class of low-density hot Jupiters with weak or featureless transmission spectra, most likely shaped by the presence of high-altitude clouds. Although a highly metal-enriched atmosphere cannot be fully excluded, current evidence disfavours this scenario. Further observations with next-generation facilities will be essential to characterise the atmospheric properties of this challenging target.
\end{abstract}
  

%
%-------------------------------------------------------------------
\section{Introduction} \label{sec:intro}
The discovery and characterisation of low-density transiting exoplanets has significantly expanded our understanding of planetary structure, formation, and atmospheric evolution. These objects, typically with densities substantially lower than that of Jupiter ($\sim$ 1.33 g$\,$cm$^{-3}$) and often comparable to or lower than those of Saturn ($\sim$ 0.69 g$\,$cm$^{-3}$) and water (1 g$\,$cm$^{-3}$), challenge classical models of planet internal structure and formation. 
Such low-density planets are predominantly found among hot Jupiters and sub-Saturns, with radii significantly inflated relative to their masses, suggesting the presence of extended gaseous envelopes dominated by hydrogen and helium \citep{Laughlin_2011, Fortney_2010, Baraffe_2014}.
Their inflated atmospheres 
are believed to be the result of complex processes, including intense stellar irradiation, residual heat from formation, and, in some cases, tidal and ohmic heating \citep{Guillot_2002, Batygin_2011, Thorngren_2018}.
Low-density exoplanets are particularly valuable targets for atmospheric studies. Their large-scale heights, combined with favourable transit geometries, enhance the amplitude of spectroscopic signatures during transit, making them ideal laboratories for probing atmospheric composition, dynamics, and cloud formation processes. 

Over the past two decades, atmospheric characterization has transitioned from the first statistical detections to detailed chemical and dynamical inventories. The field was pioneered by space-based observations at low-to-medium spectral resolution, notably the first detection of an exoplanetary atmosphere (sodium) in the inflated hot Jupiter HD 209458 b \citep{Charbonneau_2002}.
This milestone was followed by the development of ground-based High-Resolution Spectroscopy (HRS), both in transmission and emission. By resolving narrow planetary line cores and exploiting the Doppler shift of the planet along its orbit, HRS allows the planetary signal to be separated from stellar and telluric features. The first high-resolution detections of alkali metals were reported for the low-density targets HD 189733 b and HD 209458 b \citep{Redfield_2008, Snellen_2008}, targets that continue to serve as benchmark laboratories, with recent transmission \citep[e.g.][]{Blain_2024_HD189, Blain_2024_HD209} and emission HRS studies \citep[e.g.][]{Finnerty_2024, Finnerty_2025}. 

High-resolution techniques soon enabled landmark identifications of molecular species, starting with the first detection of carbon monoxide \citep{Snellen_2010} and followed by water vapour \citep{Birkby_2013, Brogi_2016}. 
Since these pioneering works, complementary observations from both space- and ground-based facilities have revealed a wide range of atmospheric properties in low-density and inflated exoplanets, including absorption features \citep{Palle_2017, Nikolov_2018, Changeat_2020, Giacobbe2021, Mounzer_2022} as well as signatures of clouds and hazes at high altitudes \citep{Line_2013, Crossfield_2013, Libby-Roberts_2020, Samra_2023, Sicilia_survey}. Furthermore, both transmission and emission spectroscopy have provided unique insights into atmospheric dynamics, including planetary rotation, heat redistribution, and high-altitude winds \citep{Louden_2015, Seidel_wind, Ehrenreich_2020, Murphy_2025}, while more recent observations and modelling studies have further demonstrated the potential of HRS to probe atmospheric circulation and wind structures \citep{Savel_2023, Zhang_2026, Seidel_2025_vertical}.

In some cases, low-density exoplanets have shown evidence of atmospheric escape in the form of extended helium tails observed during transit. The detection of near-infrared absorption from metastable helium at $\sim$1083.3 nm, as observed in WASP-107\,b and HAT-P-67\,b, indicates that these planets are undergoing significant mass loss driven by stellar irradiation \citep{Allart2018,Nortmann2018, Gully2023, Vissapragada_2024, Sicilia-hatp67}. The morphology of the helium signal, often asymmetric and spatially extended, points to preferential escape from the dayside hemisphere and suggests the presence of a leading or trailing tail shaped by stellar wind interactions. These observations provide direct evidence of ongoing hydrodynamic escape, further underscoring the importance of low-density planets as key probes of atmospheric evolution in extreme stellar environments.

Despite the progress made over the past decade, the diversity within the population of low-density exoplanets remains poorly understood.
A substantial fraction of these objects have yet to be explored from an atmospheric perspective, and each new target provides a valuable opportunity to test models of planetary formation, evolution, and atmospheric dynamics in highly irradiated, low-gravity environments. 

From a formation standpoint, the existence of extremely low-density planets raises fundamental questions about the efficiency of core accretion, the timing and extent of gas envelope accumulation, and the role of stellar irradiation and photo-evaporation in shaping long-term atmospheric evolution. Some of these planets may have formed beyond the snow line and migrated inward \citep{Mordasini_2012}. Alternatively, others may owe their low densities to internal inflation mechanisms and delayed atmospheric cooling. These internal processes are often more relevant than substantial mass loss, which is generally inefficient for relatively massive gas giants
\citep{Baraffe_2014}.

In this context, we present the first atmospheric investigation at high resolution of the gas giant WASP-13\,b \citep{Wasp13-2009}, a low-density ($\rho<0.2$ g\,cm$^{-3}$) transiting exoplanet orbiting a G1V star ($V=10.4$\,mag)
of approximately solar mass, with a period of $\sim$ 4.4 days. Discovered by the SuperWASP-North survey, WASP-13\,b belongs to the population of inflated hot Jupiters, with a radius significantly larger than expected for its mass. This combination of low surface gravity and relatively high equilibrium temperature ($g$ = 5.44 $\pm$ 0.73 m\,s$^{-2}$, \Teq \, = 1531 $\pm$ 37 K, \citealt{Southworth_2012}), calculated assuming a Bond albedo of zero and full day–night heat redistribution, implies a large atmospheric scale height, estimated to be on the order of 800 km, making it a particularly favourable target for transmission spectroscopy.

%Despite these favourable properties
Despite this, WASP-13\,b has not been the subject of any prior atmospheric studies. %and its atmospheric composition, thermal structure, and cloud properties remain unconstrained. 
Previous investigations have primarily focused on its orbital and physical parameters \citep[e.g.][]{Southworth_2012, Barros_2012, Bonomo_2017, Saha_2024}, 
the measurement of the sky-projected spin–orbit alignment via the \ac{RM} effect \citep{Brothwell_2014}, and the characterisation of the host star \citep[e.g.][]{Fossati-wasp13, Biazzo_2022}. One photometric secondary eclipse measurement at 4.5 $\mu m$ was reported by \citet{Kilpatrick_2017} using Spitzer/IRAC, providing a constraint on the planet’s dayside brightness temperature. However, no spectroscopic atmospheric studies of WASP‑13\,b have been published to date, and its atmospheric composition, temperature profile, and cloud properties remain entirely unconstrained.

In this work, we present new spectroscopic observations of WASP-13\,b aimed at probing its atmospheric composition, assessing the presence of clouds, hazes or evaporation processes. Our analysis provides the first constraints on the atmospheric properties of this target and allows us to place WASP-13\,b in context with other low-density gas giants, contributing to the broader effort to understand the physical processes governing planetary inflation and atmospheric evolution.

This paper is organised as follows. In Sect. \ref{sec:obs}, we present the observations used in this work, and the basic data reduction. In Sect. \ref{sec:parameters}, we re-evaluate the system parameters, and analyse the \ac{RM} effect. The methods applied for the extraction of the transmission spectrum and the results obtained are reported 
in Sect. \ref{sec:analysis}.
In Sect. \ref{sec:disc}, we discuss and interpret our results. Finally, in Sect. \ref{sec:conclusions}, we summarise our findings and provide our conclusions.

%%%%%%%%%% TABLE with LOG OBS %%%%%%%%
\begin{table*}
\centering
\begin{threeparttable}
\caption{Summary of the GIARPS data used in this work.}\label{tab:observing_logs}
    \begin{tabular}{l |c|ccc|ccc|c c}
    \multicolumn{10}{c}{\textit{GIARPS}}\\
    \hline
    \hline
    Night & Phase coverage& $N_{\rm obs}$\tnote{\textit{a}} & $t_{\rm exp}$[s] &  S/N (min-max)\tnote{\textit{b}} & $N_{\rm obs}$\tnote{\textit{a}} & $t_{\rm exp}$[s] & S/N (min-max)\tnote{\textit{c}} & Airmass\tnote{\textit{d}} & Seeing\\
    & & \multicolumn{3}{c|}{\textit{HARPS-N}} & \multicolumn{3}{c|}{\textit{GIANO-B}} & & \\
      \hline
    N1: 2019 Dec 26 & -0.028 $\rightarrow$ +0.030 & 20 (12) & 900 & 20.6-29.9 & - &  - & - & 1.62-1.00-1.11 & 1.7" \\
    N2: 2022 Jan 4 & -0.016 $\rightarrow$ +0.031 & 20 (15) & 900 & 22.4-41.3 & 68 (49) &  200 & 12.4-22.2 & 1.10-1.00-1.27 & 1.0" \\
    N3: 2025 Jan 18 & -0.025 $\rightarrow$ +0.031 & 24 (17) & 900 & 33.2-44.3 & 99 (64) & 200 & 15.3-22.8 & 1.07-1.00-1.76 & 1.0"\\
    N4: 2025 Feb 22 & -0.027 $\rightarrow$ +0.030 &  24 (17) & 900 & 12.6-39.9 & 67 (45) & 300 & 15.5-25.8 & 1.44-1.00-1.15 & 1.4"\\
       \hline
    \end{tabular}
   \rule{0pt}{3ex}  
\begin{tablenotes}\footnotesize
\item[\textit{a}] In parentheses, the number of analysed spectra that are considered in-transit (between the first and fourth contact).
\item[\textit{b}] The \ac{SN} is reported per pixel, computed by averaging the mean S/N of each spectral order over all the HARPS-N spectral orders.
\item[\textit{c}] The \ac{SN} is reported per pixel, computed by averaging the mean S/N of each spectral order over all the GIANO-B spectral orders.
\item[\textit{d}] The values are extracted from the FITS header of the HARPS-N spectra and indicate the airmass at the beginning, during its minimum, and at the end of the observing sequence.
%\item[*] From the 25 out-of-transit exposures, 15 exposures after the transit were disregarded due to cloud contamination.
\end{tablenotes}
\end{threeparttable}
\end{table*}

\section{Observations and Data Reduction} \label{sec:obs}
For the analysis of the WASP-13\,system, we used high-resolution visible (VIS) and near infrared (NIR) spectra acquired with the GIARPS (GIANO-B + HARPS-N) observing mode \citep{Claudi_2018} at the Telescopio Nazionale Galileo (TNG), located in La Palma (Canary islands, Spain).
We observed a total of four transits of WASP-13\,b: two within the framework of the GAPS (Global Architecture of Planetary Systems, P.I. Micela; \citealt{GAPS_Covino}) large program, and two as part of BRIDGES (Building a Road to the In-Depth investiGation of Exoplanetary atmosphereS, P.I. Borsa), which continues the GAPS legacy specifically focused on atmospheric characterisation.
The GIARPS mode enables simultaneous observations in the VIS with HARPS-N (0.39 - 0.68 $\mu$m, $R \simeq$ 115,000) and in the NIR with GIANO-B (0.95 - 2.45 $\mu$m, $R \simeq$ 50,000), allowing us to investigate both the system architecture and the planetary atmosphere across a wide spectral range \citep[e.g.][]{Baratella_2020, Carleo_2020, Guilluy2020}. A summary of the observation logs is provided in Table \ref{tab:observing_logs}.

For the VIS observations acquired with \harpsn, the exposure time ($t\rm_{exp}$) was set to 900 s. One hour of in-transit coverage was lost during the first night (N1) due to telescope malfunctions, while the second night (N2) lacks the pre-transit phase, and its first spectrum was excluded due to low \ac{SN}. The remaining two transits (N3 and N4) are complete, with N3 providing the highest \ac{SN} among the four %, while N4 is the one with the lowest \ac{SN} 
(see top panel of Fig. \ref{fig:snr_vis_nir}).
Sky spectra were retrieved simultaneously with the science observations thanks to a dedicated fibre (fibre B), pointing at a fixed
position approximately 10" from the target star, ensuring the same atmospheric conditions in both spectra.

For the NIR channel, N1 was discarded because, besides the aforementioned loss of coverage of one hour occurred during the transit, the GIANO-B autoguide encountered multiple failures. In addition, the first six spectra of N2 were excluded due to extremely low \ac{SN}. Observations during N2 were acquired in nodding mode ABAB, where target and sky spectra were taken in pairs while alternating between two nodding positions
along the slit (A and B) separated by 5"; observations during N3 and N4 used the stare mode, in which the target is observed at a fixed position along the slit. The $t\rm_{exp}$ of the observations was set at 200 s (in N2 and N3) and 300 s (in N4). %, for GIANO-B. 
The \ac{SN} of the NIR observations is reported in the bottom panel of Fig. \ref{fig:snr_vis_nir}.

When performing high-resolution measurements, it is important to consider the velocity (Doppler) smearing effect. This occurs because the \ac{RV} of the planet continuously changes along its orbit during a single exposure. If this orbital motion causes the spectral lines to shift across the detector by more than one resolution element during the integration time, the recorded planetary features become artificially broadened and blurred, which can attenuate the amplitude of the signal. Considering the change in \ac{RV} during the transit of WASP-13\,b, the maximum velocity smearing expected during a single HARPS-N exposure is 2.07 \kms, corresponding to $\sim$0.8 resolution elements. For GIANO-B, the smearing reduces to $\sim$0.46 \kms ($\sim$0.08 resolution elements). While the smearing in HARPS-N is a significant fraction of a resolution element, it is an unavoidable consequence of the exposure time required to achieve a sufficient \ac{SN} for this target. To ensure the accuracy of our atmospheric constraints, we properly account for this effect in our analysis by convolving all the synthetic templates (Secs. \ref{sec:cc-analysis} and \ref{sec:disc}) with a normalised, boxcar kernel.

The standard data reduction of the raw \harpsn \, spectra, including bias subtraction, flat-fielding, and wavelength calibration, was already performed by the HARPS-N Data Reduction Software (DRS, v3.2.0; \citealt{Cosentino_2012, Cosentino_2014})\footnote{The official HARPS-N DRS user manual is available at \url{https://www.tng.iac.es/instruments/harps/data/tng-man-harpn-0006_harps-n_drs_manual_i1r3.pdf}.}, hosted at the IA2 data centre. %\footnote{\url{https://www.ia2.inaf.it}}. 
The DRS produces both 2D spectra (in the observer's reference frame) and 1D spectra (in the barycentric reference frame of the Solar System).

The GIANO-B raw spectra were extracted using the {\tt GOFIO}\footnote{\url{https://atreides.tng.iac.es/monica.rainer/gofio}} pipeline \citep{Rainer2018}, which performs standard pre-processing operations such as dark current subtraction, flat field correction, and bad pixel removal. For the nodding observations, for each nodding pair AB, we extracted signal A from A-B and signal B from -(A-B) with an optimal
extraction algorithm~\citep{horne1986}. In this way, we removed the detector noise level and the background contribution from the spectra. For spectra obtained in stare mode, we subtracted the dark calibration frame from each spectrum and the sky emission contribution. The latter was measured in each frame, since the sky was naturally observed in a position of the slit different from that of the target during the observations. In particular, once the borders of the extraction profile have been defined, if there are enough pixels (>3) between the borders of the profile and the borders of the order, then a median of those pixels is computed for each column of the order and then subtracted from the whole column. After removing the sky, the rest of the optimal extraction follows the steps described in~\citet{horne1986}. The stare mode allowed us to acquire $~\sim30$\% more in-transit spectra in N3 with respect to N2, with the same $t_{\rm exp}$, and to integrate $~\sim38$\% more planetary signal in N4 with respect to N2, with a similar number of spectra.

For our analysis we used the "ms1d" spectral products. These consist in multi-spectral 1D outputs, where the echelle orders are kept separated and the constant step in pixel is preserved (i.e., constant step of \ac{RV} equal to half of the resolution element). Each "ms1d" output is a multi-dimensional FITS bintable with 50 rows (one for each echelle orders) and 4 columns: the number of the echelle order, the wavelength (nm), the flux, and the S/N. For each observing night, a data matrix with dimension $N_{\rm pixels}\times N_{\rm obs}$, with $N_{\rm pixels} = 2048$, is associated with each of the 50 echelle orders. It is worth noting that the 50 echelle orders falling on the detector are the ones from optical order 32, which corresponds to the reddest one, to 81, which corresponds to the bluest one. For the GIANO-B data analysis, we used the GUIBRUSHR pipeline (Amadori et al., in review), which allowed us to further reduce the extracted 1D spectra (i.e., to correct the spectral drift and refine the wavelength calibration), as described in the following, and to perform all the steps of the data analysis from the telluric removal to the cross-correlation, as described in Sect.~\ref{sec:data-reduction} and~\ref{sec:cc-analysis}.

A necessary step of the data reduction consists in refining the wavelength calibration provided by GOFIO.
Indeed, GOFIO performs a preliminary wavelength calibration using an uranium-neon arc lamp in the vacuum wavelength frame. 
However, since the lamp is only observed at the end of the night to avoid persistence on the detector, the GOFIO wavelength solution is not sufficiently accurate; thus, it is expected to shift and jitter during the observations due to the mechanical instability of the instrument. To correct for the spectral drift of the instrument, for each order, we aligned all the spectra of a single night to a common reference frame via a cross-correlation with a time-averaged observed spectrum of the target used as a template. Thanks to this correction, we achieved a residual shift of the spectra of less than one-tenth of a pixel ($\lesssim 0.3$\,km\,s$^{-1}$) for most of the spectral orders. To refine the initial calibration, we adopted the approach described in~\citet{Giacobbe2021}, where telluric lines are matched between the observations and a telluric model to re-calibrate each order. For our analysis, we excluded a priori a fixed set of spectral orders dominated by tellurics (echelle orders 40-42, 54-56) and the Y band (orders 71-81) in which GIANO-B has a characteristic drop of throughput. For the single night analyses, in each night we also excluded a few orders with high residual drift, or for which the refined wavelength calibration procedure failed. When combining different nights, in order to use the same orders for the different datasets, we discarded a fixed set of orders built by the union of the orders discarded in the single nights. In the second column of Table~\ref{tab:spectral_orders_gianob}, we report the orders discarded in our analysis.

%%%%%%%%%% TABLE with system parameters %%%%%%%%
\begin{table*}
\centering  
\caption{
Stellar, planetary, and orbital parameters adopted in this work. The values for the planetary radius and mass, the semi-major axis and the planet \ac{RV} semi-amplitude were derived from \citet{Bonomo_2017}, \citet{Biazzo_2022} and \citet{Patel_2022}.}
\label{tab: parameters} 
\renewcommand{\arraystretch}{1.3}
\begin{tabular}{l l c c} \hline \hline
\noalign{\smallskip}
Parameter & Symbol [unit]  & Value & Reference \\ \hline 
\noalign{\smallskip}
\textbf{Star} & & & \\
\noalign{\smallskip}
%$m_\mathrm{V}$ (mag) & & & \\
Spectral type & Sp.T. & G1V & \cite{Biazzo_2022}  \\
Stellar mass & $M_\star$ [\Msun] & $ 1.19 \pm 0.05$ & \cite{Biazzo_2022}\\
Stellar radius & $R_\star$ [\Rsun] & $1.54 \pm 0.05$ & \cite{Biazzo_2022}\\
Stellar effective temperature & \teff [K] &  $5980 \pm 50 $ & \cite{Biazzo_2022}\\
Stellar metallicity & \feh [dex] &   $0.10 \pm 0.08 $ & \cite{Biazzo_2022}\\
Stellar surface gravity & \logg [$\log_{10}$(cm s$^{-2}$)] &  $4.09 \pm 0.12 $ & \cite{Biazzo_2022} \\
Stellar age & [Gyr] & $4.7 \pm 1.0 $ &\cite{Biazzo_2022}\\
Parallax & [mas] & $4.331 \pm 0.020 $ & \cite{GaiaDR3_correct}\\
%Distance & $d$ [pc] & 230.9 $\pm$ 1.1 & \cite{GaiaDR3_correct}\\
Distance & $d$ [pc] & 229.266$^{+0.934}_{-1.045}$ & \cite{Bailer-Jones_2021}\\
Stellar RV semi-amplitude & $K_\star$ [\ms] & $56.89^{+0.89}_{-0.93}$  & \cite{Bonomo_2017}\\ 
Projected stellar rotation velocity & \vsini [\kms] & 4.54 $\pm$ 0.12 & This work\\ 
\noalign{\smallskip}
\textbf{Planet} & & \\
\noalign{\smallskip}
Planet mass & $M_\mathrm{p}$ [\MJ]& $0.515 \pm 0.020$ & Derived\\
Planet radius & $R_\mathrm{p}$ [\RJ] & $1.430 \pm 0.049$ & Derived\\
Planet equilibrium temperature & \Teq [K]&  1531 $\pm$ 37& \cite{Southworth_2012}\\
Orbital inclination & $i_\mathrm{p}$ [deg] & $85.43 \pm 0.29$ & \cite{Brothwell_2014}\\
Orbital projected obliquity & $\lambda$ [deg] &  2.73$^{+1.08}_{-1.14}$ & This work\\
Planet RV semi-amplitude & $K_\mathrm{p}$ [\kms] & $137.7 \pm 8.2$ & Derived\\
\noalign{\smallskip}
\textbf{System} & &  \\
\noalign{\smallskip}
Orbital period & $P_\mathrm{orb}$ [days] & $4.35301101 \pm 0.00000066$ & \cite{Kokori_2023} \\
Mid-transit time & $T_\mathrm{c}$ [BJD] & $2457077.30360 \pm 0.00028$ &\cite{Kokori_2023}\\
Transit duration & $T_{14}$ [h] & $4.003 \pm 0.024$ & \cite{Brothwell_2014}\\
Semi-major axis & $a$ [AU] &  $0.0556 \pm 0.0024$ & Derived\\
Systemic radial velocity & \vsys [\kms] & $9.855 \pm 0.010 $  &\cite{Biazzo_2022}\\
% others?
\hline 
\end{tabular}

\end{table*}

\section{System parameters and spin-orbit analysis}\label{sec:parameters}
We adopted updated stellar and planetary parameters for the WASP-13\,system, based on literature values and on our own analysis. A summary of all parameters used in this work is reported in Table \ref{tab: parameters}.

We re-derived the planetary mass (\mplanet) and radius (\rplanet) starting from the stellar parameters published in \citet{Biazzo_2022}. For \rplanet, we combined their stellar radius (\rstar) with the transit depth ($\propto$ \rplanet/\rstar) reported by \citet{Patel_2022}, which provides one of the most precise measurements available to date. For \mplanet, we adopted the radial velocity semi-amplitude ($K_\star$) from \citet{Bonomo_2017}, whose determination is also among the most precise in the literature. Regarding the systemic velocity (\vsys), we chose the value derived by \citet{Biazzo_2022}, which is more precise and fully consistent with that reported in Gaia DR3 \citep{GaiaDR3_correct}, ensuring a modern and accurate velocity zero-point. All these inputs were used to produce a coherent and internally consistent set of stellar and planetary parameters for our analysis.

A first measurement of the sky-projected spin–orbit angle ($\lambda$) of WASP-13\,b was presented by \citet{Brothwell_2014}, based on SOPHIE observations acquired with the 1.93 m telescope at the Observatoire de Haute-Provence (OHP). Using \ac{RV} data from a single transit, they found that the planet is in a well-aligned prograde orbit, with a projected obliquity of $\lambda$ = 8$^\circ$$^{+13}_{-12}$. This result is consistent with the general trend of low obliquities in systems hosted by cool stars (\teff < 6250 K), as observed in statistical studies of spin–orbit alignments \citep[e.g.][]{Albrecht_2012}.

\begin{figure}
    \centering
    \includegraphics[width=\linewidth]{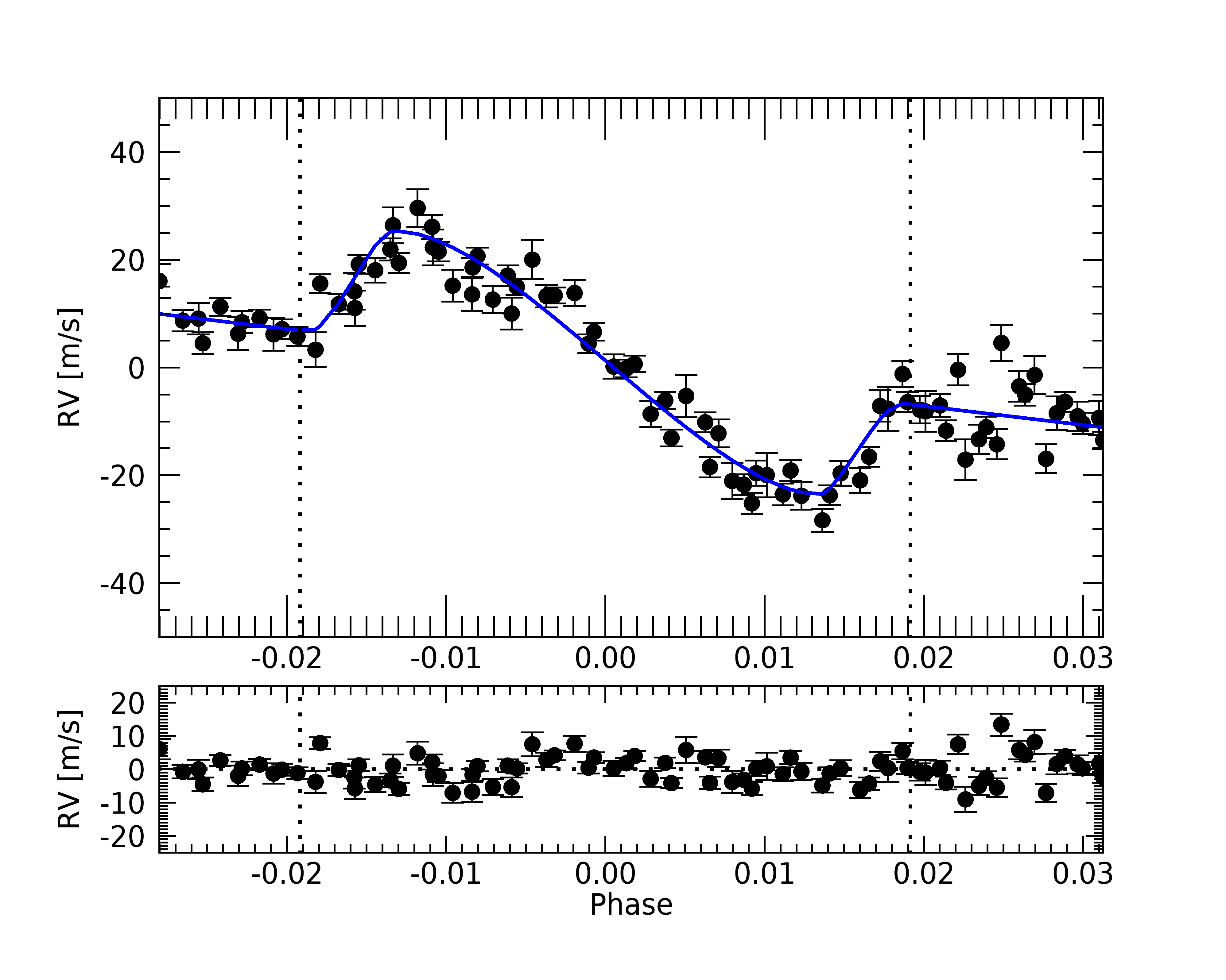}
    \caption{Rossiter-McLaughlin effect. Radial velocities time series taken during the four transits and shifted with respect to the \vsys. The best-fit model is superimposed, and the corresponding residuals are shown in the lower panel. The dotted black lines indicate the points of the first and fourth contact of the transit.
    }
    \label{fig:rm_analysis}
\end{figure}

With the availability of new high-resolution spectroscopic observations from GIARPS, covering multiple transits, we revisited the measurement of the \ac{RM} effect to refine the determination of both $\lambda$ and the projected stellar rotational velocity (\vsini). The \ac{RM} effect is visible in our dataset, both in the two-dimensional \ac{CCF} maps and more clearly in the \ac{RV} time series (Fig. \ref{fig:rm_analysis}). Given the moderate \ac{SN} of some of the spectra and the relatively low \vsini, we opted to model the \ac{RM} signal via classical \ac{RV} fitting rather than using Doppler shadow techniques. Doppler tomography is typically more effective for fast-rotating stars (\vsini $\gg$ 5 \kms), where the planet’s shadow produces a clear, moving distortion across the line profile that can be modelled directly \citep[e.g.][]{CollierCameron_2009, Brown_2016}.

We performed a combined fit of the \ac{RV} time series of the four observed transits (see Tables \ref{tab:rvs-rhk} - \ref{tab:rvs-rhk4}, second and third columns).
The \ac{RV} measurements were obtained from the DRS products through the 
YABI platform \citep{Hunter2012},
using a G2 mask template and a \ac{CCF} width of 40 \kms, with a step of 0.25 \kms.
In the fit, we left free to vary $\lambda$ and \vsini, which were set to be the same for all the transits. In addition, we fitted a constant \ac{RV} offset ($v_{\rm off}$) for each transit, which we report as relative shifts with respect to \vsys. This approach is mathematically equivalent to re-determining the effective systemic velocity directly from the HARPS-N stellar RVs for each individual transit ($v_{\rm sys, night} = v_{\rm sys} + v_{\rm off}$). The inclusion of these nightly offsets allows us to account for any instrumental or zero-point differences between our dataset and the literature, and ensures that the stellar reference frame is self-consistently anchored to our observations, preventing potential systemic velocity offsets from propagating into the planetary rest-frame analysis. All other system parameters, including the systemic velocity \vsys, were fixed to the values reported in Table \ref{tab: parameters}.

We derived a value of $\lambda$ = 2.7$\pm 1.1^\circ$ and \vsini = 4.5$\pm0.1$ \kms, indicating a well-aligned planetary orbit and significantly reducing the uncertainties of the previous measurement by \citet{Brothwell_2014}. 
The improved accuracy in both parameters benefits from the higher cadence and the increase in the number of observed transits in our dataset.
The velocity shifts measured for each transit are as follows: $v_{\rm off}=$10.4 $\pm$ 0.6 \ms (N1), 5.57$^{+0.56}_{-0.58}$ \ms (N2), 7.89$^{+0.43}_{-0.46}$ \ms (N3), and 22.8 $\pm$ 0.5 \ms (N4). Although the presence of an additional planetary companion could explain these offset values, 
the available data (only four sparse measurements spanning more than five years) do not allow us to investigate this scenario further or place meaningful constraints on the properties of a potential companion; the observed offsets are at the level of a few tens of \ms and therefore cannot be uniquely attributed to a specific orbital configuration.  

We inspected the \logrhk chromospheric activity index for each night (fourth and fifth columns of Tables \ref{tab:rvs-rhk} - \ref{tab:rvs-rhk4} and Fig. \ref{fig:rhk}). We extracted it from the HARPS-N spectra through the YABI platform, adopting a \textit{B-V} colour index of 0.79 $\pm$ 0.10 mag, as derived from SIMBAD photometry, although the YABI implementation requires a fixed \textit{B–V} value and does not account for its uncertainty. This value appears somewhat redder than expected for a G1V star \citep{Pecaut-Mamajek}, which may reflect photometric uncertainties and/or a small contribution from interstellar reddening; nevertheless, this has no impact on our results, as the derived \logrhk values remain uniformly low. 
The average values measured for the four transits (-5.153$\pm$0.002, -5.099$\pm$0.002, -5.085$\pm$0.001, -5.121$\pm$0.005, for N1, N2, N3, and N4, respectively) show no significant variation, supporting the hypothesis of a quiescent chromosphere \citep{Fossati_2015}. We note that the final part of the N4 time series exhibits larger uncertainties, which could partly explain the anomalous velocity shift observed that night.
A visual inspection of the \ion{Ca}{II} H \& K line cores (Fig. \ref{fig:caii_hk}) reinforces the conclusion that WASP-13 is a quiet star, as the profiles remain stable across all nights.
For completeness, we also inspected the average out-of-transit CCF profiles for each observing night  (Fig. \ref{fig:ccf_out}). The profiles are consistent across the different nights and show no significant residuals with respect to the master out-of-transit CCF, indicating no evidence for variability in the stellar line profiles.

\section{Atmospheric characterisation}\label{sec:analysis}
\subsection{Stellar and telluric contamination removal}\label{sec:data-reduction}
For the HARPS-N data, following the same approach adopted for HAT-P-67 b in \citet{Sicilia-hatp67}, we extracted the transmission spectrum for each night of observation and the combined one
from the 2D spectra using the {\tt SLOPpy}\footnote{\url{https://github.com/DanielaSloppy/SLOPpy}} (Spectral Lines Of Planets with python) public pipeline \citep{Sicilia}. 
While the pipeline automatically subtracts the sky spectrum acquired through fibre B from the science spectrum, thereby correcting for sky background and possible contamination from telluric emission features,
the removal of telluric absorption lines is performed with version 3.8 of {\tt Molecfit} \citep{Smette_2015, Kausch_2015}, which is wrapped within {\tt SLOPpy} through a system call.  The stellar models used for the correction of the \ac{CLV} and \ac{RM} effects, which are the two main processes altering the transmission spectra, have been obtained with the software package 
{\tt PySME} \citep{PySME}; we used the line list from the third version of Vienna Atomic Line Database (VALD3\footnote{\url{https://vald.astro.uu.se}}) 
together with ATLAS9 stellar atmosphere models \citep{Atlas9}, adopting non-LTE grids for the relevant atomic species. Following the approach described in \citet{Sicilia}, synthetic stellar spectra were computed at different limb angles by generating a set of spectra at 21 values of $\mu = \cos\theta$ (from 0.001 to 1). The stellar disk was then discretized into surface elements using a regular square grid with a fixed cell size of $0.01R_\star \times 0.01 R_\star$. Each surface cell is associated with the corresponding limb-angle spectrum interpolated from the computed $\mu$ values. To account for the \ac{RM} effect, the spectrum of each surface element was Doppler-shifted according to the projected stellar rotational velocity and the geometry of the system.
For each exposure obtained during transit, the stellar spectrum blocked by the planet was computed by integrating the spectra of the occulted surface elements and subtracting them from the disk-integrated stellar spectrum. The resulting phase-dependent stellar model was used to derive a wavelength-dependent correction factor, which was applied to the individual transmission spectra by dividing the observed spectra by the corresponding CLV and RM correction
before combining them in the planetary reference frame.

For the NIR analysis, telluric (stationary) and stellar (quasi-stationary) lines were removed by performing a principal component analysis (PCA), as it is often done in high-resolution analysis in the NIR (e.g., ~\citealt{dekok2013}). For each observing night, we ran PCA on each spectral order separately (i.e. on each $N_{\rm pixels}\times N_{\rm obs}$ matrix), after having expressed the data in the logarithmic space, where most of the systematics are additive rather than multiplicative. In order to run PCA, we pre-processed the data following the method described in~\citet{basilicata2025} to mask the spectral channels strongly affected by telluric contamination and to standardise the data. In particular, we normalised each spectrum to its median value to correct the baseline flux variations. Then, we masked the spectral channels with stronger or saturated telluric lines. To do this, we subtracted the time-averaged spectrum to all the spectra (i.e. we subtracted to each spectral channel its mean value), we computed the standard deviation of data in each spectral channel and its median value ($\sigma_{\rm m}$), and then we masked the spectral channels that had a standard deviation $\sigma > 1.5\cdot \sigma_{\rm m}$. After subtracting the time-averaged spectrum from all the spectra and masking highly contaminated spectral channels, we reduced the data to a null-mean variable by subtracting the mean value computed on the different spectral channels from each spectrum. Then, we followed the method suggested by Spring \& Birkby (in review) and adopted by~\citet{basilicata2025} to select the appropriate number of principal components to remove from the data. In particular, for the selection of the number of components, we considered the relative decrease of the standard deviation of the residuals ($\Delta\sigma=\frac{\sigma_{i-1}-\sigma_i}{\sigma_{i-1}}$, with $i$= the $i$-th number of principal components removed) with the increase of the number of components removed, and studied its relative variation ($\delta=\frac{\Delta\sigma_{i}-\Delta\sigma_{i-1}}{\Delta\sigma_{i}}$) to see when a plateau was reached. We then placed a threshold at $-0.1$ and selected the minimum number of principal components for which $-0.1 < \delta < 0$ for all the spectral orders. In Fig.~\ref{pca_components_plot}, for each observing night, we report the variation of the standard deviation of the residuals, used to select the number of principal components removed in the analysis. According to this method, we removed  12 principal components for N2, 7 principal components for N3, and 5 principal components for N4. Finally, for each night, we reconstructed the selected number of components following the procedure described in detail by \citet{Giacobbe2021}, and removed them to all the spectral orders. After the telluric correction, we applied a smoothing filter with a kernel size of 40 pixels to suppress low-frequency trends in the spectra, and we exponentiated the residual matrix associated to each spectral order to express back data in the linear space.

\begin{figure*}
    \centering  
    \includegraphics[trim=0 0 0 25,clip,width=0.3\linewidth]{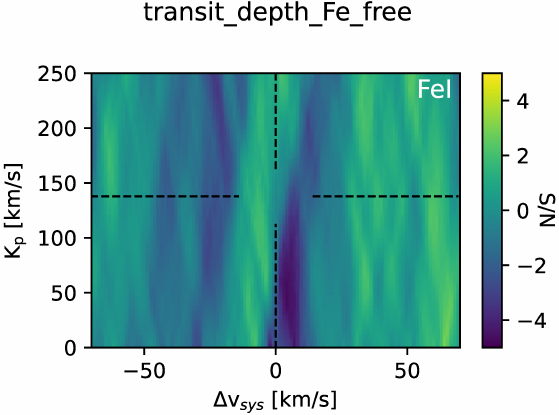}
    \includegraphics[trim=0 0 0 25,clip,width=0.3\linewidth]
    {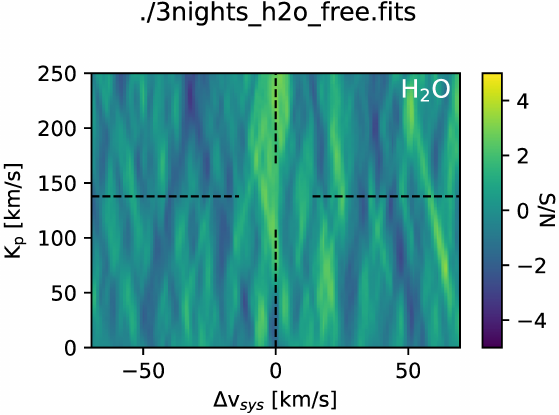}
    \includegraphics[trim=0 0 0 25,clip,width=0.3\linewidth]
    {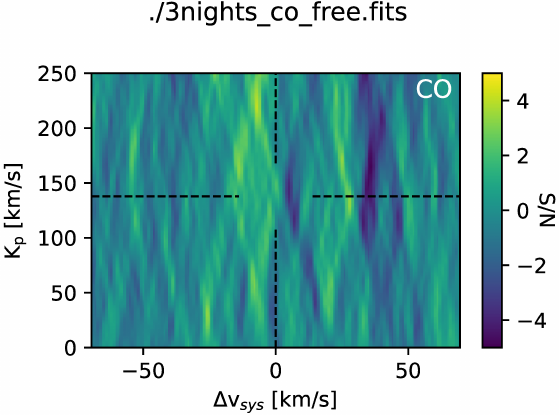}
    \caption{Planes of K$\rm_p$-$\Delta$\vsys \, obtained by combining all nights for the \ion{Fe}{I}, H$_2$O, and CO templates. For each species, the corresponding template has been computed assuming free chemistry with \ac{VMR}=10$^{-3}$. In each plot, the dashed black cross indicates the expected K$\rm_p$ and $\Delta$\vsys \, of the system.}
    \label{fig:combined_Kp-vsys}
\end{figure*}

\subsection{Cross-correlation analysis}\label{sec:cc-analysis}
We performed a cross-correlation analysis on the transmission spectra in both the VIS and NIR ranges, extracted as described in Sect. \ref{sec:data-reduction}. For both wavelength domains, we employed the same set of forward models covering the full range from 0.35 to 2.45 $\mu m$. The models were generated with \texttt{petitRADTRANS} v3, assuming an isothermal profile at $T= 1531$\,K, a pressure range from 10$^{-8}$ to 10 bar with 70 layers, and a reference pressure of 0.1 bar. No clouds, H$^-$ opacity, or additional continuum sources beyond H$_2$-H$_2$ and H$_2$-He collision-induced absorption were included, and solar abundances were adopted as baseline. Two classes of single-species models were produced: free-chemistry models, with a constant-with-altitude \ac{VMR} of 10$^{-3}$ for the investigated species; chemical equilibrium models, where the \ac{VMR} of each species was computed using the chemical network \texttt{Chemcat}\footnote{\url{https://chemcat.readthedocs.io/en/main/chemistry_tutorial.html}} (Cubillos et al. in prep.) for solar metallicity and C/O ratio. It is worth noting that in chemical equilibrium the \acp{VMR} depend on the temperature and pressure profile and even an
isothermal profile, as the one we chose, can lead to a non-constant \acp{VMR}.

We investigated the presence of multiple chemical species that are common in the atmosphere of hot giant planets, including \ion{Ca}{I}, \ion{Cr}{I}, \ion{Mg}{I}, \ion{Fe}{I}, \citep{Kurucz_2005}, \ion{K}{I}, \ion{Na}{I} \citep{Burrows2003},  H$_2$O (POKAZATEL, \citealt{pokazatel2018}), and CO (HITEMP, \citealt{hitemp2010}). In the NIR, we first searched for H$_2$O and CO, since at the equilibrium temperature of WASP-13\,b, assuming a solar composition, H$_2$O and CO are expected to be the most abundant secondary molecular species in the atmosphere of this kind of target~\citep{madhusudhan2012,moses2014}, and thus are expected to produce a stronger absorption signal.
Each single-species model was cross-correlated with the data over a grid of systemic (\vsys) and planetary velocities (i.e. different planetary radial-velocity semi-amplitudes $K_{\rm p}$). Before cross-correlation, all model spectra were convolved to the instrumental resolution to ensure a direct comparison with the data. We searched for the atomic species using the HARPS-N data, while for the molecular species we used GIANO-B data, as these species have most of their absorption features in the VIS and NIR, respectively.

\begin{figure*}
    \centering
    \includegraphics[trim=0 0 0 25,clip,width=0.3\linewidth]{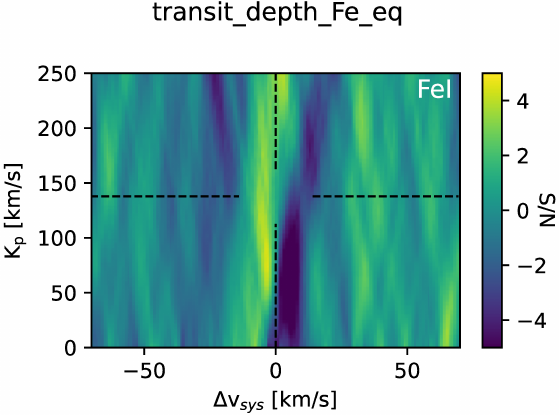}
    \includegraphics[trim=0 0 0 25,clip,width=0.3\linewidth]{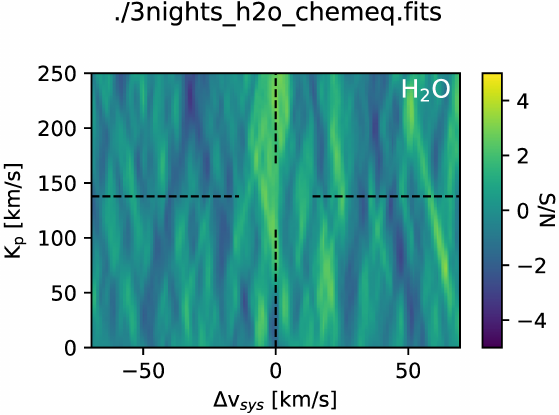}
     \includegraphics[trim=0 0 0 25,clip,width=0.3\linewidth]{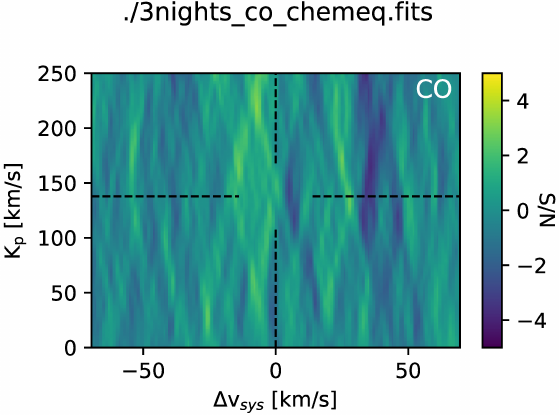}
    \caption{Same as Fig. \ref{fig:combined_Kp-vsys} but assuming chemical equilibrium.}\label{fig:combined_Kp-vsys_equi}
\end{figure*}

By combining the data of the four nights (excluding N1 for GIANO-B), no signals attributable to the planet's transmission spectrum were found in the VIS or NIR range for any of the species considered. Fig.~\ref{fig:combined_Kp-vsys} shows the results obtained with the free-chemistry models for the \ion{Fe}{I}, H$_2$O and CO templates, but we note that the chemical equilibrium case yielded consistent conclusions (see Fig. \ref{fig:combined_Kp-vsys_equi}). Since most of the CO absorption is concentrated at %$2400$\,nm 
$\sim$ 2.4 $\mathrm{\mu}$m
(the strongest band), for this molecule we also performed the cross-correlation analysis by using only the K band of GIANO-B spectral range, where the strongest CO absorption band falls. In the third column of Table~\ref{tab:spectral_orders_gianob}, we report the orders used to search for CO in this restricted band. Even in this case, by combining the nights, we do not obtain any significant signal at the expected planetary radial velocity (see the left and middle panels of Fig.~\ref{fig:map_Kp-vsys_co_order1}).

\begin{figure*}
    \centering
    \includegraphics[width=0.4\linewidth]{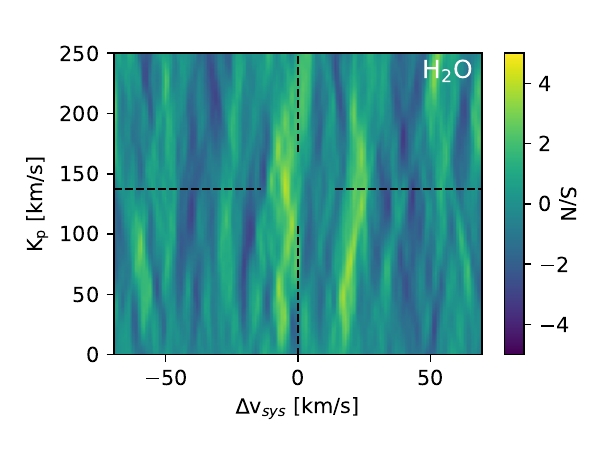}
    \includegraphics[width=0.4\linewidth]{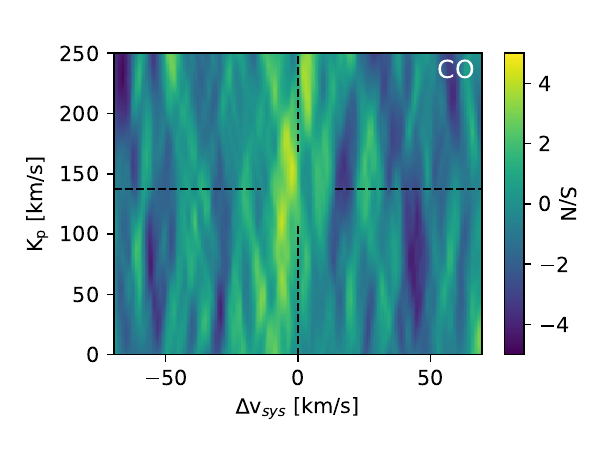}\\
    \includegraphics[width=0.45\linewidth]{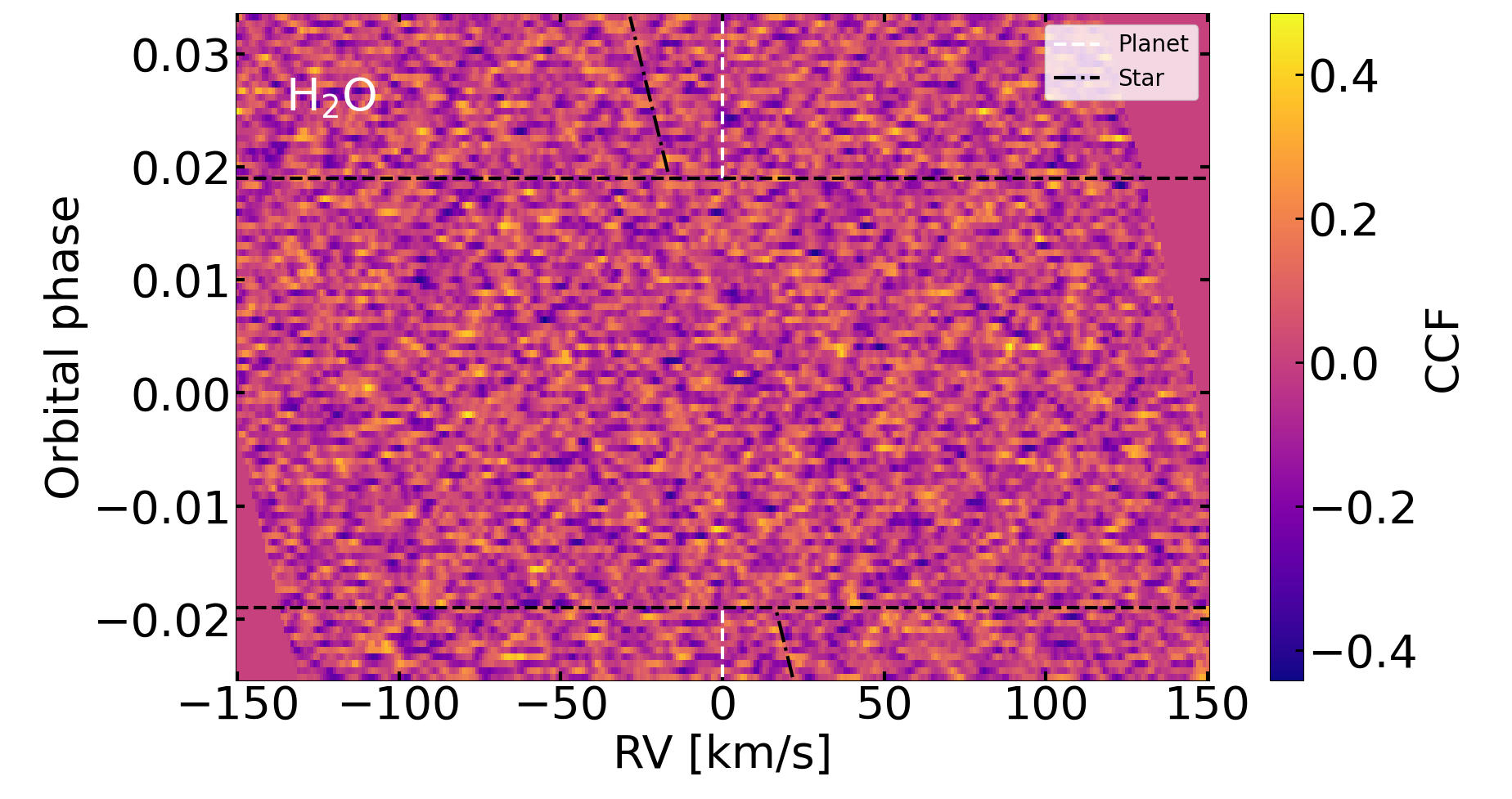}
    \includegraphics[width=0.45\linewidth]{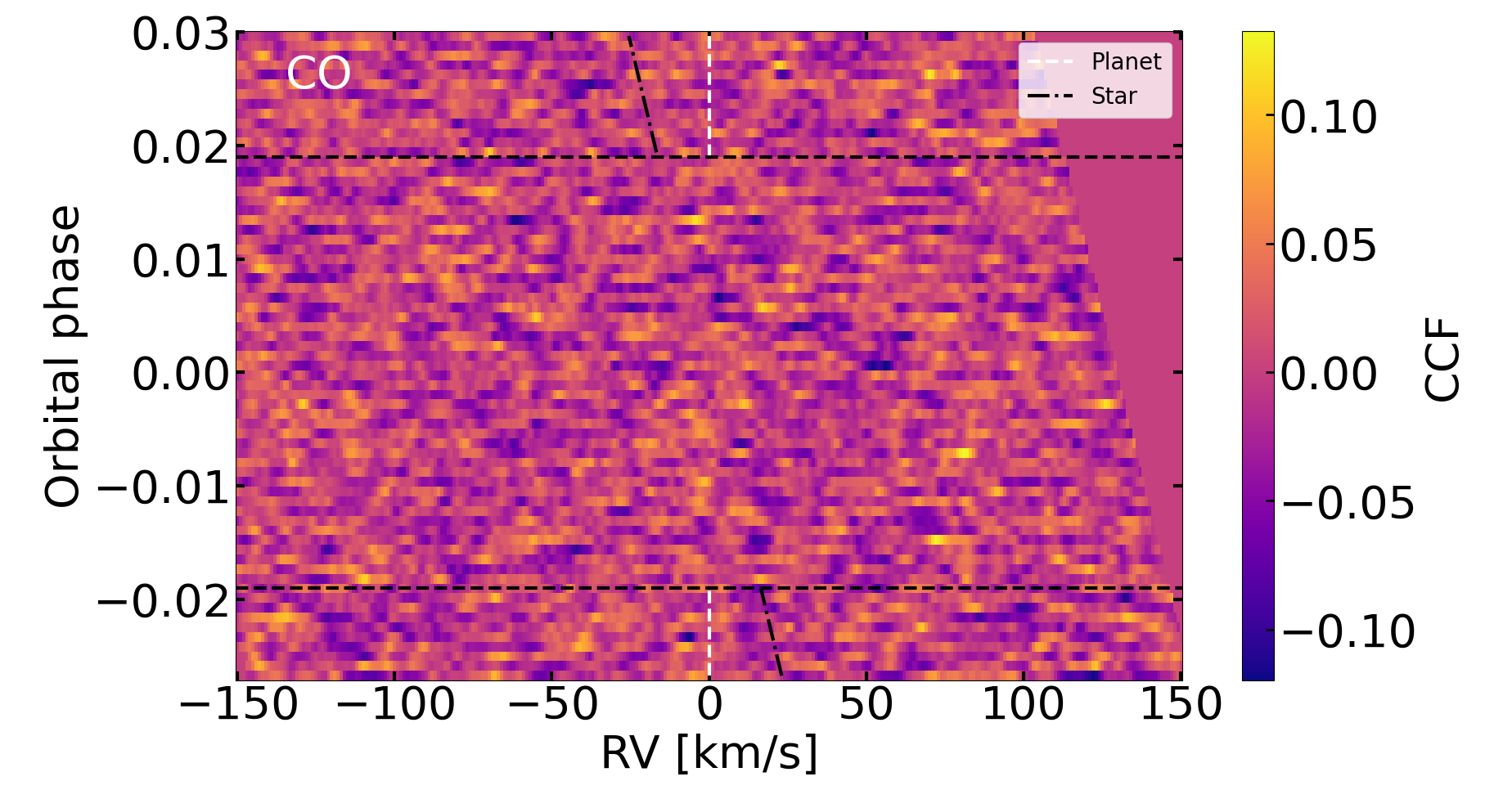}
    
    \caption{Cross-correlation maps obtained by using the dataset of N3 (left panels) and N4 (right panels), for the H$_2$O and CO templates (using only the K band of GIANO-B for the CO), respectively. \textit{Top}: Planes of K$\rm_p$-$\Delta$\vsys, in which a peak of cross-correlation is visible close to the expected planetary RV for both species but the signals are not sufficiently robust to be claimed as detections (see the text for more details). \textit{Bottom}: Cross-correlation as a function of the orbital phase in the planet rest-frame. The dashed white (dash-dotted black) line represents the planet (stellar) RV.}
    \label{fig:map_Kp-vsys_tentative}
\end{figure*}

It is worth noting that in N3 and N4, we measure a tentative signal of H$_2$O and CO (when using only the K band), respectively, around the expected planetary RV (Fig.~\ref{fig:map_Kp-vsys_tentative}). These signals does not seem to be due to a spurious cross-correlation with telluric residuals, as we repeated the analysis by not considering the contribution to the CCFs from the frames of N3 and N4 for which the \ac{RV} of the signal was $\leq10$\,km\,s$^{-1}$ from the telluric one, and we were still able to measure the two tentative signals, with a smaller \ac{SN}, as expected from the smaller number of spectra used. However, when dealing with possible planet CO signals, it is important to consider possible stellar contaminations. Indeed, in the photosphere of G stars, like WASP-13, or cooler spectral types, CO is present and produces absorption bands in the stellar spectrum that could cross-correlate with planet atmospheric CO templates, producing spurious cross-correlation signals that affect the CCF analysis and, in the worst case, can be misinterpreted as planet features. The fact that we do not measure any residual CO signal at the stellar RV in our main analysis proves that PCA is effective in removing stellar CO lines. However, \ac{RM} and \ac{CLV} effects cause line distortions not corrected by PCA, leaving possible stellar residuals in our data that could still affect the planet signal. We quantified the magnitude of these effects, in order to evaluate the potential impact on our result. In particular, the average depth of stellar CO lines for this type of star, whose depth is mainly influenced by the temperature of the photosphere, is of 0.04 (4\%) with respect to the continuum level (measured on a high-resolution stellar spectrum from the G\"{o}ttingen Spectral Library {\tt PHOENIX\footnote{\url{https://phoenix.astro.physik.uni-goettingen.de/?page_id=15}}}). The RM effect would introduce a distortion of the stellar line to the level of $(R_{\rm p}/R_{\star})^2 = 0.009$ (0.9\%), for a total change in the stellar CO line depth of $0.04\cdot0.009 = 3.6\cdot10^{-4} = 0.036\%$. Since the average depth of planetary CO lines is $0.1\%$ (measured from the free-chemistry CO atmospheric template), the amplitude of these effects is comparable to the amplitude of the planet CO signal. Therefore, we decided to perform a more accurate correction of the stellar spectrum to asses the actual impact of these effects on the CO tentative signal. In particular, we computed synthetic spectra from 3D simulations of stellar convection resolved both spatially and temporally, and we
coupled them with a model reproducing the geometry of the transit of WASP-13\,b, following the methods described in~\citet{Chiavassa-Brogi}, using the stellar and planet parameters adopted in this work. In this way, for each orbital phase of each night, we had a realistic stellar spectrum distorted by RM and CLV effects due to the transiting planet. For each orbital phase, we continuum-normalised the corresponding stellar model Doppler-shifted at the stellar \ac{RV}, and divided the GIANO-B data covering the K band by it, before the PCA step of the analysis. Repeating the CCF analysis with the CO templates on the star-corrected spectra and combining the signal of the three nights, considering only the K band, we do not measure any significant difference in our results (see the right panel of Fig.~\ref{fig:map_Kp-vsys_co_order1}), confirming the minor impact of these effects in our dataset and the absence of a clear CO detection.

For this reason, and due to the low significance of the tentative CCF signals and to the fact that these signals disappear when combining multiple nights, we do not claim any detection for H$_2$O and CO and conclude that we do not robustly measure the atmospheric signal of WASP-13\,b neither through atomic nor molecular absorption features with cross correlation.

\subsection{Search for chemical species through single line analysis}
In addition to the cross-correlation approach, we performed a single-line analysis of selected atomic species. This complementary method allows us to directly investigate strong individual lines that are expected to probe different regions of the planetary atmosphere and provide constraints on distinct physical processes. In particular, we focused on the metastable helium triplet (\ion{He}{I}), the H-alpha line (H$\alpha$) and the sodium doublet (\ion{Na}{I} D), which have been extensively discussed in the context of atmospheric escape and upper-atmosphere dynamics.

\begin{figure}
    \includegraphics[width=\linewidth]{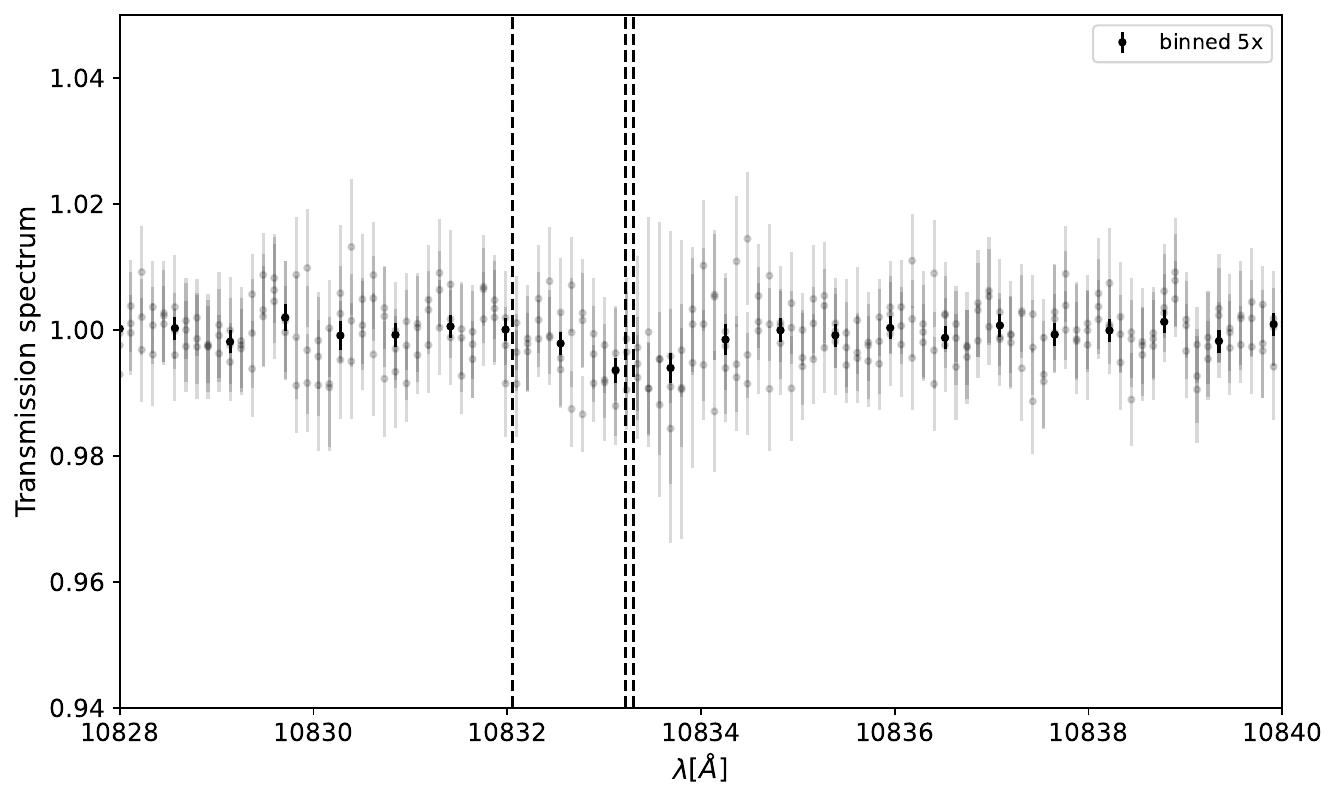}
    \includegraphics[width=.97\linewidth]{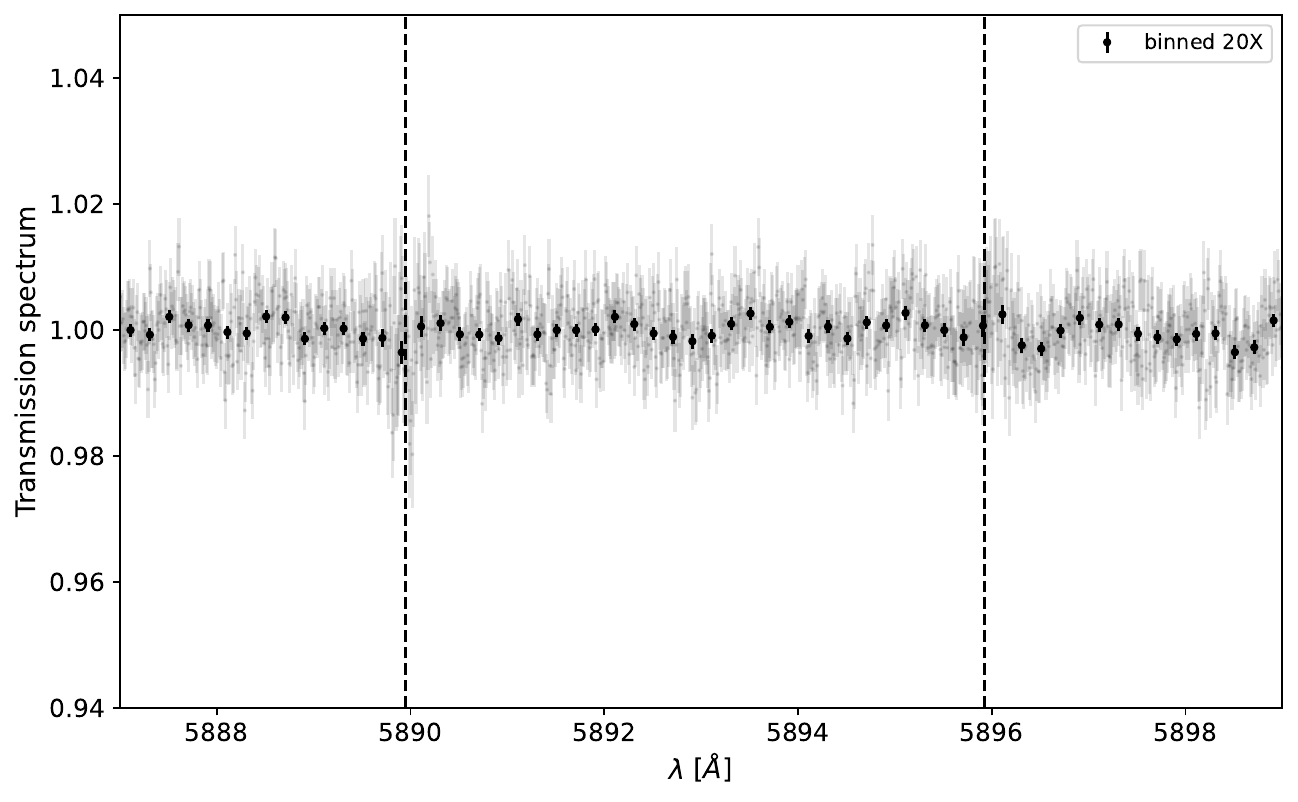}
    \caption{Final transmission spectra of WASP-13\,b centred around the \ion{He}{I} triplet (upper panel), and the \ion{Na}{I} doublet (bottom panel), whose rest frame transition wavelength is indicated by the vertical dashed lines. The choice of the binning factor is driven by the different number of available data points.}
    \label{fig:ts_He-Na}
\end{figure}

\begin{figure}
\includegraphics[width=0.9\linewidth]{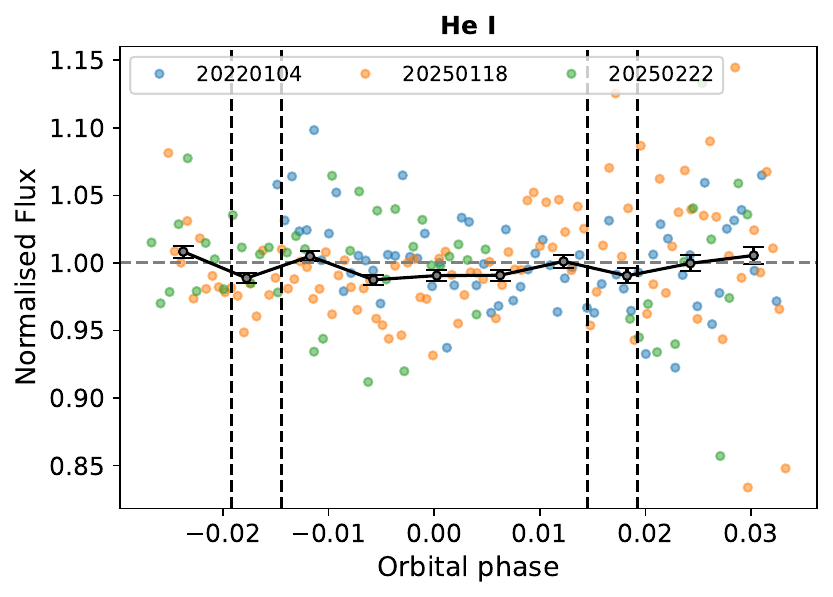}
\includegraphics[width=0.9\linewidth]{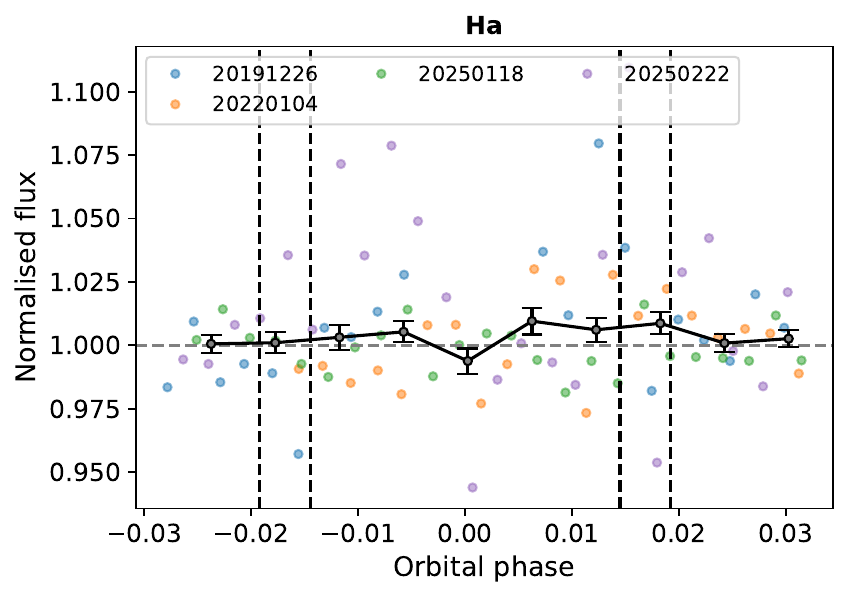}
\includegraphics[width=0.9\linewidth]{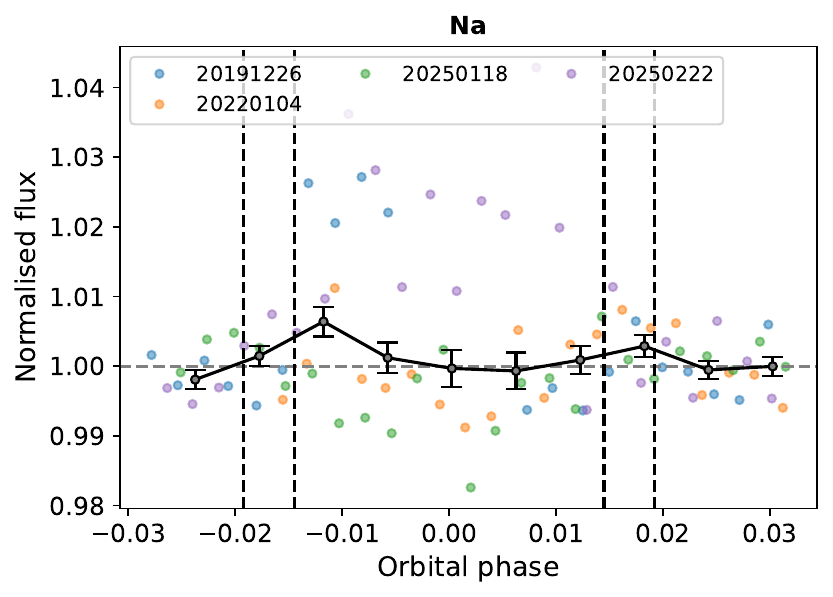}
\caption{Spectroscopic light curves in the planet rest frame computed in a bandpass of 0.75 \AA \,around \ion{He}{I} (upper panel), and 0.40 \AA \,around H$\alpha$ and \ion{Na}{I} (middle and bottom panels).
Black points are computed with a phase bin of 0.003. Vertical black dashed lines indicate the position of the transit contact points t$_1$, t$_2$, t$_3$, t$_4$. The error bars on the \ion{He}{I} light curve are smaller compared to those of \ion{Na}{I} and H$\alpha$. This difference is primarily attributed to the higher number of data points available in each phase bin for \ion{He}{I}.}
\label{fig:LCs}
\end{figure}

\begin{figure*}
    \centering
    \includegraphics[width=\textwidth]{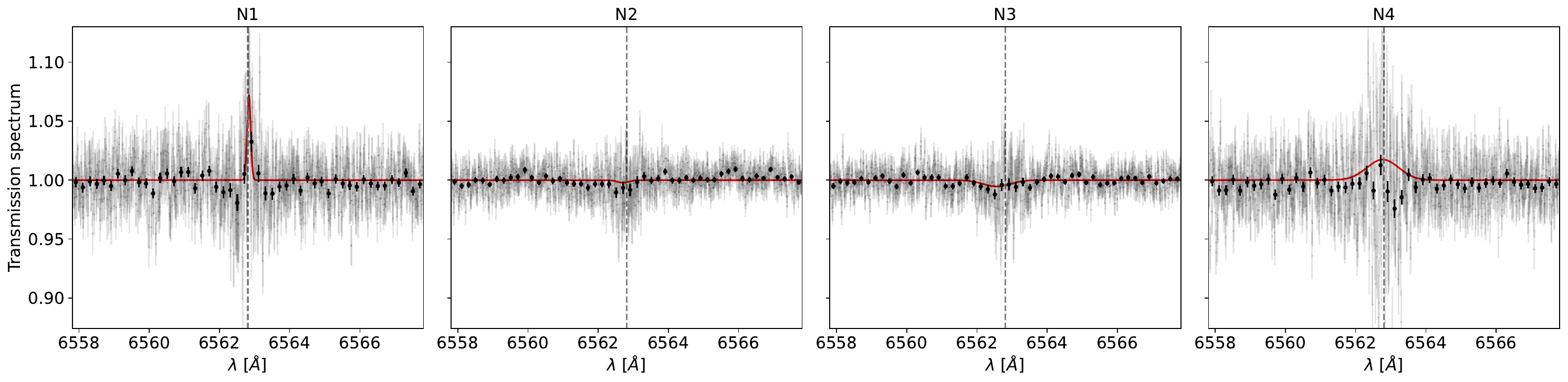}
    \caption{Transmission spectra of WASP-13\,b for each night. The spectra are centred around the H$\alpha$ line in the planetary rest frame (light grey) and also binned 20 times (in black circles). The red line is the MCMC Gaussian fit performed by \texttt{SLOPpy}, while the vertical dashed line indicates the rest frame transition wavelength of the H$\alpha$ line.}\label{fig:ts_Halpha}
\end{figure*}

A full description of the methodology applied to the \ion{He}{I} can be found in \citet{Guilluy2024}, \citet{Sicilia-hatp67}, and \citet{Darpa_2024}; here, we provide an overview of the main steps.
 From the full set of echelle orders, we selected order number 71 for the subsequent analysis, as it contains the \ion{He}{I} triplet of interest.
Telluric absorption was corrected using the ESO {\tt Molecfit} software \citep{Smette_2015, Kausch_2015}, with special attention to water vapour lines near 1083.51~nm. The strong OH emission line around 1083.43 nm was masked following the procedure outlined in \citet{Guilluy2023}. The next step involved isolating the planetary signal via transmission spectroscopy. First, the spectra were shifted to the stellar rest frame and normalised to unity via median division, excluding the spectral region containing the helium triplet. A master out-of-transit spectrum, $S_\mathrm{out}(\lambda)$, was constructed by averaging spectra acquired outside the transit window (i.e. phases earlier than first contact $t_1$ or later than fourth contact $t_4$). 
To investigate the possible presence of an evaporating or escaping \ion{He}{I} atmosphere, individual transmission spectra $T(\lambda, i)$ were computed by dividing each in-transit spectrum by $S_\mathrm{out}(\lambda)$, and then interpolated in the planet’s rest frame. The upper panels of Fig. \ref{fig:ts_He-Na} and Fig.~\ref{fig:LCs} display the average transmission spectrum and the spectroscopic light curve, respectively, %\citep[][]{Allart2019},
while Fig.~\ref{mappe_Hei} shows the 2d transmission spectra maps in the planet's rest frame for each investigated night. 
The light curve was obtained by integrating the transmission spectrum within a central passband of 0.75 \AA\, around the stellar \ion{He}{I} triplet (centered at $\sim$10833.3 \AA). In this spectral region, the definition of reliable continuum regions is complicated by the presence of telluric residuals, silicon features, and fringing patterns. While these effects can be mitigated in the average transmission spectrum, they cannot be robustly corrected at the level of individual spectra used to construct the light curve. Therefore, consistently with previous works \citep[e.g.][]{Allart2019}, we derive the \ion{He}{I} spectroscopic light curve using only the central passband and no reference continuum bands.
We did not find evidence of any planetary \ion{He}{I} absorption. 
We quantified the upper limits on the non-detection by fitting a Gaussian profile to each in-transit mean transmission spectrum (computed over $t_2$–$t_3$) with a differential evolution Markov chain Monte Carlo (DE-MCMC) framework \citep{TerBraak2006}. Correlated noise was modelled using Gaussian Process (GP) regression within the same DE-MCMC routine, employing a squared exponential kernel as the covariance function \citep{Guilluy2024, Sicilia-hatp67}. An additional white noise (jitter) component was included.
%$\sigma_\mathrm{j}$. 
We put a 2$\sigma$ upper limit of 3.1 $\%$ on the \ion{He}{I} absorption, corresponding to an effective radius of $R_{\rm eff} =$ 2.15 $R_{\rm p}$ (1$\sigma$ = 1.2~$\%$, $R_{\rm eff} =$ 1.55 $R_{\rm p}$).

The H$\alpha$ line is another valuable diagnostic of the upper atmospheric layers, tracing excited hydrogen that can be produced by stellar irradiation and escape processes. We searched for the H$\alpha$ signal in our optical transmission spectra extracted from \texttt{SLOPpy}, by summing all in-transit observations divided by the master out in the planet's reference system. The spectroscopic light curve is computed by integrating the signal within a central passband of 0.40 \AA \, around the line core at 6562.8 \AA, and normalizing it using reference continuum bands at [6558.0 - 6561.0] \AA \, and [6564.0 - 6567.0] \AA. The result for each night is shown in Fig. \ref{fig:ts_Halpha}, while the spectroscopic light curve obtained by combining all nights is visible in Fig. \ref{fig:LCs} (middle panel). The analysis revealed apparent emission features in N1 and N4, and weak absorption in N2 and N3. 
We performed an MCMC fit for each of the four observed transits, adopting as free parameters the line contrast, FWHM, velocity shift, K$\rm_p$ (with a prior centred at 140 km/s), and an instrumental jitter term. The fits yield apparent non-zero contrasts in three out of four nights, with values of +7.2$^{+8.5}_{-2.6}\%$ (N1), –0.5$^{+0.3}_{-0.5}\%$ (N3), and +1.7$^{+5.1}_{-0.6}\%$ (N4), while one night (N2) remains consistent with zero (–0.2$^{+0.9}_{-1.1}\%$). Although the 68 $\%$ credible intervals for N1, N3, and N4 formally exclude zero, the large uncertainties, the alternating emission and absorption features, and their lack of repeatability across transits indicate that the observed variability is more likely due to stellar surface inhomogeneities or statistical noise than to a planetary signal. We therefore conclude that no robust and repeatable detection of H$\alpha$ can be claimed from our dataset, with any potential planetary absorption constrained to be weaker than a few percent in relative contrast.

While the \ion{Na}{I} D is primarily a tracer of the upper atmospheric layers and has been widely used to study atmospheric composition, pressure broadening, and cloud coverage, in some cases it has also been proposed as an indirect indicator of atmospheric escape. This is particularly the case when broad absorption features or high-velocity wings are observed, which may trace the presence of extended or escaping atmospheric layers driven by hydrodynamic outflows (e.g. \citealt{Seidel_2020}).
We fitted the \ion{Na}{I} D in the transmission spectra of the four transits using a joint MCMC analysis for the D$_2$ and D$_1$ lines with shared FWHM and velocity shift. The MCMC fits of both lines generally yield flat posterior distributions, with either vanishingly small or unrealistically broad line widths, indicating the absence of coherent planetary absorption. The only exception is N4, where only the D$_2$ line shows an emission-like feature 
that does not necessarily correspond to physical processes, but can arise from residual stellar effects or imperfect removal of \ac{CLV} and \ac{RM} signals. The fact that this feature is more pronounced when the spectra are aligned in the stellar rest frame, and that it is not consistently detected in both components of the \ion{Na}{I} doublet, strongly disfavours a planetary origin. This interpretation is further supported by the behaviour of the stellar activity index \logrhk, which remains constant on average across all epochs but shows larger uncertainties during N4, especially toward the end of the transit (see Table~\ref{tab:rvs-rhk4}). These enhanced uncertainties are likely driven by the lower S/N of the spectra in that phase, although short-term stellar variability cannot be completely ruled out.
When combining all four nights, the joint fit yields contrasts consistent with zero for both components of the \ion{Na}{I} D and no absorption is visible either in the final transmission spectrum (Fig. \ref{fig:ts_He-Na}, bottom panel) or in the spectroscopic light curve (Fig. \ref{fig:LCs}, bottom panel). 
The latter was computed using a central passband of 0.40 \AA \, around each line of the doublet (D$_2$ at 5889.95 \AA \, and D$_1$ at 5895.92 \AA), and reference continuum bands at [5874.89 - 5886.89] \AA \, and [5898.89 - 5910.89] \AA. 
Overall, we attribute the apparent D$_2$ variability to a combination of limited \ac{SN} and stellar contamination, and we report non-detections of sodium in the transmission spectra. Consequently, we do not measure the atmospheric absorption signal of WASP-13\,b with single-line analysis.

\begin{figure*}
    \centering
    \includegraphics[trim=0 0 60 27,clip,width=.35\linewidth]{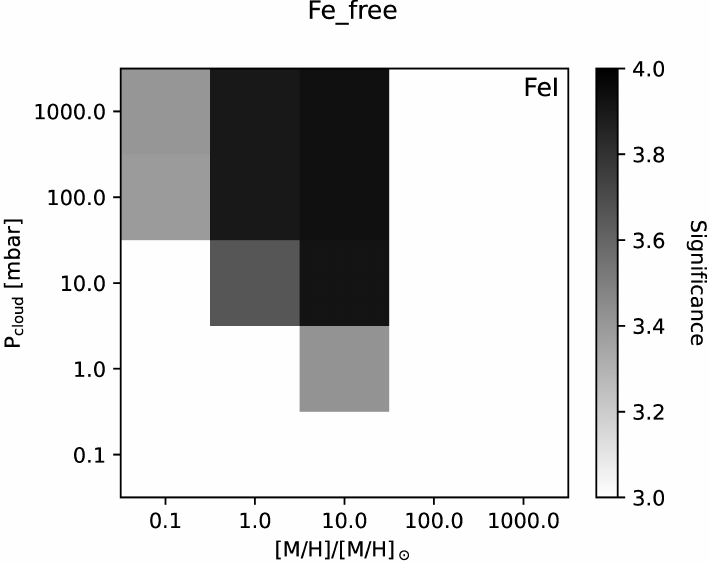}
    \hspace{1mm}
    \includegraphics[trim=0 0 60 27,clip,width=.35\linewidth]{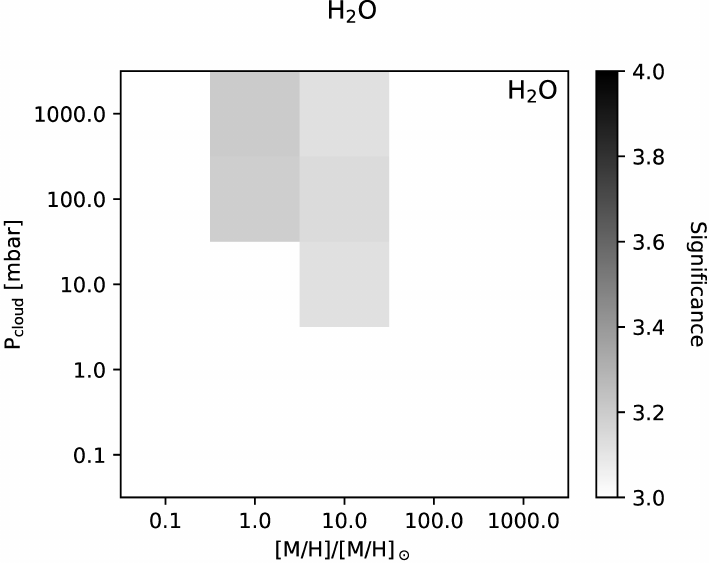}
    \hspace{1mm}
    \includegraphics[trim=280 0 0 27,clip,width=.075\linewidth]{injectionTestsH2O.pdf}
    \caption{Maps of detectability for the \ion{Fe}{I} (left panel) and H$_2$O (right panel) injection models. Each box represents an atmospheric configuration spanning a range of metallicities ([M/H]) and cloud deck altitudes ($P_{\rm cloud}$). A common colour scale is adopted and intentionally limited to \ac{SN} $\lesssim$ 4 to preserve the visibility of the weaker H$_2$O signals. The maximum recovered significance for \ion{Fe}{I} is \ac{SN} $\simeq$ 3.9.}
    \label{fig:detectability}
\end{figure*}

\section{Discussion} \label{sec:disc}
The lack of a measurable atmospheric signal in the VIS and NIR, with both cross-correlation and single-line analysis techniques, could indicate the presence of an atmosphere with muted spectroscopic features, or that the S/N of our dataset is not sufficiently high to probe the atmosphere of this target. We also noted that during the analysed nights, the planet \ac{RV} overlaps with the telluric one for most of the transit duration (in N4 it is not properly overlapped, but it is at $\leq 10$\,km\,s$^{-1}$ in the first part of the transit). This means that if the PCA procedure applied in the NIR was too aggressive, it could have eroded the possible planetary signal.

Before exploring possible atmospheric configuration that could lead to a non-detection with our dataset, first we assessed the impact of the number of principal components removed on the results of the cross-correlation analysis. In order to be less conservative, for each night, we repeated the PCA procedure selecting a different number of components for the different spectral orders, instead of removing the same number of components for all the orders. In particular, we repeated the steps described in Sect. \ref{sec:data-reduction} but, this time, to each spectral order, we removed the minimum number of principal components for which the threshold was reached for that particular order. In this way, we removed an order-dependent number of principal components, which is less than or equal to the number of components removed in the main analysis. In particular, for N2 we removed a number of components $4\leq N_{\rm comp}\leq7$ among the different spectral orders, $4\leq N_{\rm comp}\leq5$ for N3, and $4\leq N_{\rm comp}\leq5$ for N4. By combining the three GIANO-B datasets we still do not obtain any significant signal at the expected planetary \ac{RV} (Fig. \ref{fig:3nights_free_tanlim}).

In order to evaluate our ability to detect a possible atmospheric signal with our dataset, we performed a simple set of injection-and-recovery tests.
In particular, we injected a grid of atmospheric models into the data and, for each injection, we tried to recover the injected signal by applying the same methods used to search for the real planetary signal. We built a grid of 25 models, which could be representative of the atmosphere of the planet, including the absorption of all the chemical species we probed in this work, under the following assumptions: 1D isothermal atmosphere, chemical equilibrium, and solar C/O ratio. We explored a large range of possible metallicities and cloud deck altitudes, spaced evenly on a log-scale. In particular, the metallicity could vary from 0.1$\times$ to 1000$\times$ solar metallicity ([M/H]$_\odot$) and the pressure at the top of the grey cloud deck ($P_{\rm cloud}$) could vary from 10$^{-4}$ to 1\,bar. For each model, we performed the following steps. Before the injection, we normalised the model to the continuum and, for each night, shifted it at the planetary \ac{RV} in the observer rest-frame, for each orbital phase, using the orbital parameters reported in Table~\ref{tab: parameters}. For each night, we then injected the shifted model only in the data corresponding to in-transit orbital phases (between $t_1$ and $t_4$), before the telluric removal stage of the analysis. After the injection, we repeated all the steps of our analysis on the injected data, using the same number of principal components (in the NIR) and the same spectral orders used in the main analysis. Finally, we cross-correlated the injected data with the different single-species templates adopted in the main analysis and built the injected $K_{\rm p}-\Delta v_{\rm sys}$ maps for each species (CCF$_{\rm inj}$). To remove the random noise structure at the injected position in the $K_{\rm p}-\Delta v_{\rm sys}$ maps, we subtracted to each of these map the corresponding $K_{\rm p}-\Delta v_{\rm sys}$ map obtained without the injected signal (i.e., the one obtained with real observations, CCF$_{\rm obs}$). We computed this difference by using the $K_{\rm p}-\Delta v_{\rm sys}$ maps not normalised to the corresponding noise levels. In this way, we directly computed the difference between the CCF signals in the maps to isolate the CCF signal corresponding to the injected template ($\Delta$CCF$=$CCF$_{\rm inj}-$CCF$_{\rm obs}$). Finally, we used the standard deviation of non-injected $K_{\rm p}-\Delta v_{\rm sys}$ map to convert the signal strength into units of \ac{SN}, and searched for a peak of cross-correlation at the planetary RV with a \ac{SN} $>3$, after having combined the dataset of all the nights (four for the VIS, and three for the NIR). The \ac{SN} of the peak of cross-correlation was evaluated as the maximum \ac{SN} in the interval $110\leq K_{\rm p}$ [km\,s$^{-1}$] $\leq160$ and $-10\leq \Delta v_{\rm sys}$ [km\,s$^{-1}$] $\leq10$. In this way, for each probed chemical species, we could build a map of detectability for the different injected models, which tells us for which assumed atmospheric configuration we would have been able to detect the different chemical species, with our dataset and our methods.

In Fig. \ref{fig:detectability}, we report the maps of detectability for \ion{Fe}{I} and H$_2$O. For the other probed chemical species, we were unable to recover any of the injected models with a \ac{SN} $>3$. 
For \ion{Fe}{I}, injection tests yield a maximum recovered significance of \ac{SN} $\simeq$ 3.9, allowing us to exclude some possible atmospheric configurations that would have produced a detectable signal at this level. 
As for H$_2$O, although we are able to recover only a few models with S/N $\gtrsim3$, the atmospheric configurations excluded by the test with H$_2$O are a subset of those excluded by the test with \ion{Fe}{I}, confirming the overall consistency between the two results.

We also repeated the injection and recovery tests by injecting the synthetic signal at the expected $K_{\rm p}$ but with an offset in $v_{\rm sys}$ equal to $\Delta v_{\rm sys} = -60$\,km\,s$^{-1}$. For these tests, we chose this position in the $K_{\rm p}-\Delta v_{\rm sys}$ map as it seemed relatively free from spurious cross-correlation peaks (at $\geq 3\,\sigma$), and directly evaluated the \ac{SN} on the injected $K_{\rm p}-\Delta v_{\rm sys}$ maps (i.e., without computing the $\Delta$CCF). These tests yields consistent detectability results and further confirms the robustness of the atmospheric constraints that we place.

In particular, we can exclude an atmosphere with metallicity [M/H] $\lesssim10\times$ [M/H]$_\odot$ and simultaneously a cloud top pressure $P_{\rm cloud}\gtrsim10$\,mbar. If the atmospheric metallicity is $\simeq10\times$ [M/H]$_\odot$, we would have been able to recover the signal of \ion{Fe}{I} quite regardless of the pressure of the cloud deck, probably due to the strong line contrast, so an atmospheric metallicity $\simeq10\times$ [M/H]$_\odot$ is very unlikely. In general, we can see that we cannot exclude an atmosphere with either a high metallicity ($\gtrsim100\times$ [M/H]$_\odot$) or a high-altitude cloud deck ($P_{\rm cloud}\lesssim1$\,mbar), since both configurations have the effect of reducing the line contrast, thus reducing the atmospheric signal and possibly causing our non-detections. However, a highly metal-enriched atmosphere would appear difficult to reconcile with the planet’s large radius. In the framework of inflation models, increasing the atmospheric metallicity tends to increase the mean molecular weight and opacity (e.g. \citealt{Guillot_2006, Fortney_2007}), thereby hampering radius inflation at a given level of stellar irradiation. Moreover, recent 3D cloud formation models for hot Jupiters with temperatures comparable to that of WASP-13\,b (e.g. \citealt{Helling_2023}) predict that condensates can extend to very low pressures at the terminator (below $\sim$10$^{-4}$ bar), effectively masking molecular features in transmission. In this context, the absence of detectable atomic and molecular species in our spectra would more naturally point to the presence of a high-altitude cloud deck rather than to an intrinsically metal-rich atmosphere.
If such high-altitude clouds are indeed present, the planetary radius measured from the transit would correspond to the altitude (or pressure level) of the cloud tops, rather than to that of a clear atmosphere at $\sim$0.1-1 bar. Consequently, the observed transit radius may be overestimated relative to the true radius of the optically thick atmosphere, potentially contributing to the apparently large inflation inferred for WASP-13\,b (e.g. \citealt{Lammer_2016}).

Our ability to recover a smaller number of models for H$_2$O, could be due to the fact that in the NIR we used only three nights, instead of the four used in the VIS, and that the S/N of GIANO-B dataset is smaller than that of HARPS-N (an average S/N $\sim30$ for the HARPS-N dataset and $\sim19$ for the GIANO-B dataset). Although the fact that we are not able to recover the injected signal for the other chemical species could point towards a poor quality of our dataset, the test with \ion{Fe}{I} disfavours this hypothesis as it points towards an atmosphere with either a high (super-stellar) metallicity or a high-altitude cloud deck. Nevertheless, our results underline the need to acquire more data to characterise the atmospheric properties of this target or to put stronger constraints on the possible atmospheric configurations.

Finally, it is worth noting that our considerations are based on the assumptions that we made about the atmosphere of WASP-13\,b, while the real atmospheric configuration could deviate from them. However, the models used for these tests constitute the simplest framework we could adopt to put the first reasonable constraints about the atmosphere of this difficult target.

\section{Conclusions}\label{sec:conclusions}
In this work, we present the first atmospheric study of the low-density hot Jupiter WASP-13\,b, combining four high-resolution transits observed with GIARPS. 

Our updated analysis of the \ac{RM} effect provides new system parameters, with \vsini = 4.5$\pm0.1$ \kms, and a projected obliquity of $\lambda$ = 2.7$\pm1.1^\circ$, confirming a well-aligned orbit. This result agrees with the observed trend that hot Jupiters orbiting cool stars tend to have low spin–orbit misalignments, supporting the role of tidal realignment in these systems.

From the atmospheric perspective, we performed a cross-correlation analysis across both the visible and near infrared wavelength ranges. Despite exploring a wide parameter space in systemic and planetary velocity, %the cross-correlation maps reveal no significant detection of any of the species investigated in either the optical or infrared domain.
no statistically significant detections were found for any of the atomic or molecular species considered. Although in the near infrared, tentative signals of H$_2$O and CO were identified in individual nights, they are not confirmed in the combined datasets and are not sufficiently robust, therefore more data are needed to confirm the presence of these species in the atmosphere of the target.

We also performed a single-line analysis of \ion{He}{I}, H$\alpha$, and \ion{Na}{I} to probe possible tracers of extended or escaping atmospheres. None of these species shows significant in-transit absorption. The lack of detectable Na I or He I features suggests the absence of a strong high-altitude thermosphere or ongoing atmospheric evaporation. A marginal variability in the H$\alpha$ region and in the D$_2$ line was observed; however, their lack of repeatability from night to night, the alternation between emission- and absorption-like signatures, and the large uncertainties associated with the lowest \ac{SN} phases indicate that this variability is most likely driven by statistical noise and residual stellar contamination, rather than by a genuine planetary atmospheric signal. This is supported by the observed stability of the \ion{Ca}{II} H \& K line cores and the \logrhk	index, which rule out significant chromospheric variability or intense active regions that could introduce spurious signals in the transmission spectra.
It is worth noting that the lack of single-line signals is consistent with the results from the cross-correlation analysis and reinforces the conclusion that no atmospheric features are robustly detected within our dataset.

Injection–recovery tests, performed for both atomic and molecular templates, allowed us to place the first observational constraints on the atmospheric properties of WASP-13\,b, indicating that any potential spectroscopic features are below the current detection threshold.

From an interpretative perspective, both high atmospheric metallicity and high-altitude clouds could suppress spectral features in transmission. However, a highly metal-enriched atmosphere is difficult to reconcile with the planet’s inflated radius, whereas 3D cloud-formation models predict that condensates at the terminator may reach very low pressures, naturally muting molecular lines. In this context, a high-altitude cloud deck provides a more plausible explanation for our non-detections.

Although current data do not yield a confirmed atmospheric detection, WASP-13\,b adds an important case to the population of inflated, low-density exoplanets whose energy balance and atmospheric composition remain poorly understood. 
In this context, the absence of detectable signals provides meaningful constraints on the atmospheric properties of the planet, ruling out scenarios that would produce stronger features at the sensitivity of our data. Such constraints are valuable both for guiding future observations of this target and for improving the statistical understanding of the atmospheric diversity of hot Jupiters.

Future observations with high-resolution facilities (e.g. ANDES/ELT) and with low-resolution space missions such as Ariel will be essential to characterise the atmospheric structure of this challenging planet and to place it within the broader context of the formation and evolution of low-density exoplanets.

%--------

\section*{Acknowledgements}
The authors acknowledge financial contribution from 
INAF GO Large Grant 2023 GAPS-2, from the INAF GO Large Grant 2024 "Strenghtening pylons for BRIDGES" as well as from the European Union - Next Generation EU RRF M4C2 1.1 PRIN MUR 2022 project 2022CERJ49 (ESPLORA). They also acknowledge the Italian centre for Astronomical Archives (IA2, https://www.ia2.inaf.it), part of the Italian National Institute for Astrophysics (INAF), for providing technical assistance, services and supporting activities of the GAPS collaboration. D.S. and G.S. acknowledges the funding support from CHEOPS ASI-INAF Agreement n. 2019-29-HH.0 and Addendum n. 2019-29-HH.1-2022.
M.B. acknowledges the financial support from the PRIN MUR 2022 NOCTIS project (CUP: D53C24004270006).
C.D.M. acknowledges support from the INAF Mini-Grant ``Impact of planetary Masses and Radii Estimates on the Atmospheric Retrievals – IMaREA'', awarded under the INAF Fundamental Astrophysics funding scheme. Part of the research activities presented in this paper were carried out with the contribution of the Next Generation EU funds within the Italian National Recovery and Resilience Plan (PNRR), Mission 4 – Education and Research, Component 2 – From Research to Business (M4C2), Investment Line 3.1 – Strengthening and Creation of Research Infrastructures, Project IR0000034 ``STILES – Strengthening the Italian Leadership in ELT and SKA''.
L.M. acknowledges financial contribution from PRIN MUR 2022 project 2022J4H55R.
G.M. acknowledges support by the Space It Up project funded by the Italian Space Agency, ASI, and the Ministry of University and Research, MUR, under contract n. 2024-5-E.0 - CUP n. I53D24000060005. 
This work has made use of the VALD database, operated at Uppsala University, the Institute of Astronomy RAS in Moscow, and the University of Vienna. \\
D.S. and M.B. thank M. Rainer for discussions and insights. 

\section*{Data Availability}
Data generated at a large-scale facility.

\bibliographystyle{mnras}
\bibliography{sample} 

\appendix
%\begin{appendix}
\section{Additional figures and tables} \label{add}

\begin{figure}
    \centering
    \includegraphics[width=\linewidth]{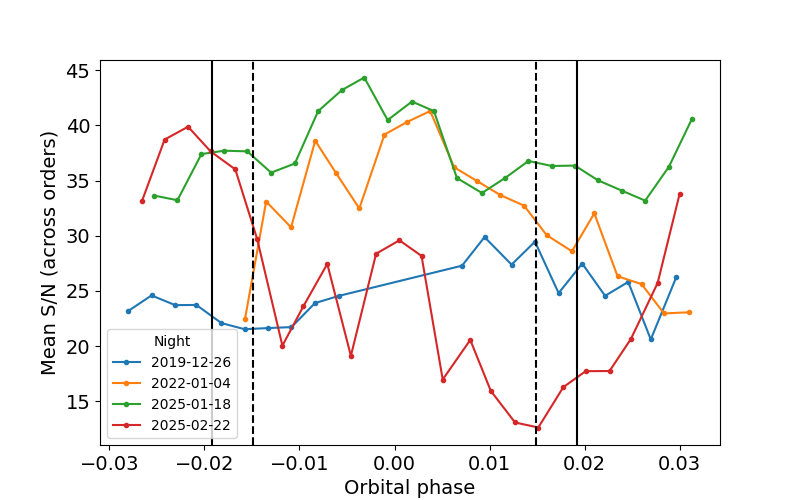}
    \includegraphics[width=\linewidth]{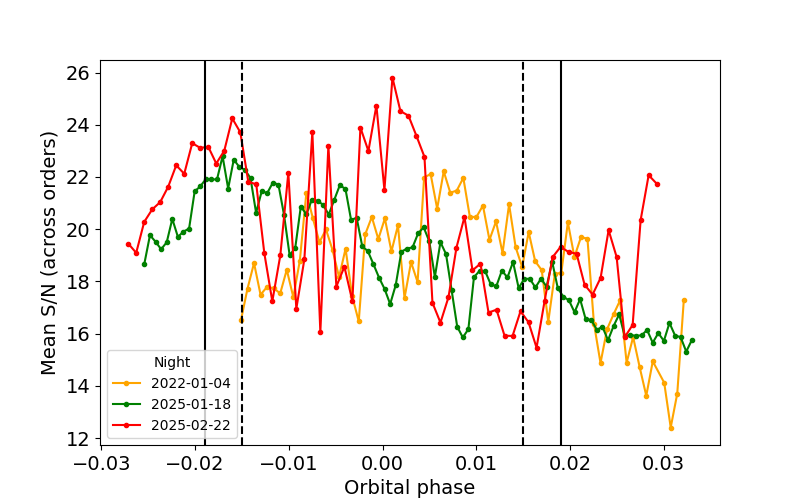}
    \caption{Mean \ac{SN} across all the spectral orders extracted from the FITS header of the HARPS-N spectra (top panel) and from the {\tt GOFIO} pipeline for the GIANO-B spectra (bottom panel) for each night. Vertical continuous lines indicate the points of first and fourth contact of the transit, while vertical dashed lines represent the points of second and third contact.}
    \label{fig:snr_vis_nir}
\end{figure}

\begin{figure*}
    \centering
    \includegraphics[width=\linewidth]{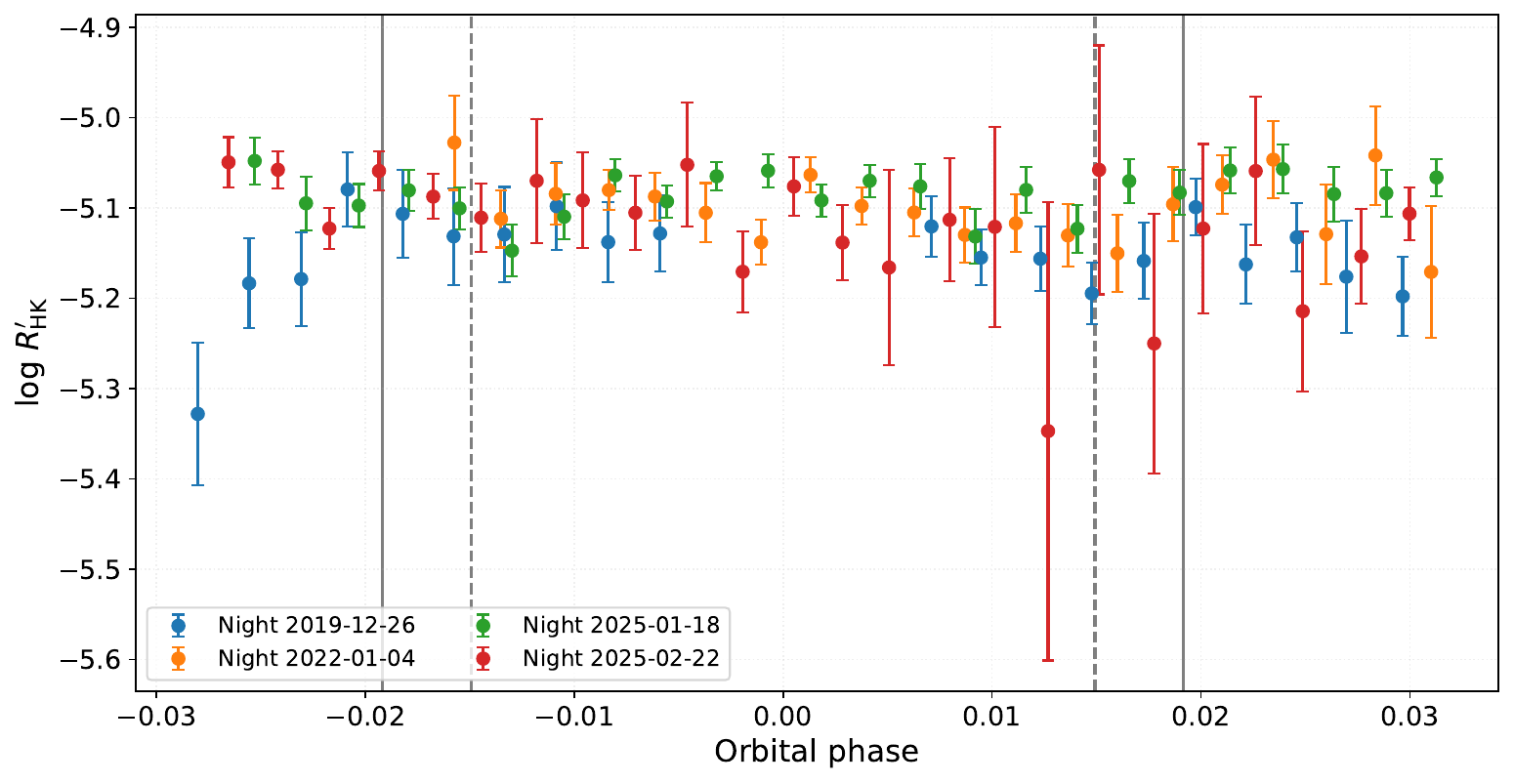}
    \caption{Activity indicator \logrhk \, as a function of the orbital phase for each night. The continuous vertical lines indicate the points of first and fourth contact of the transit, while the dashed vertical lines represent the points of second and third contact.}
    \label{fig:rhk}
\end{figure*}

\begin{figure}
    \centering
    \includegraphics[width=\linewidth]{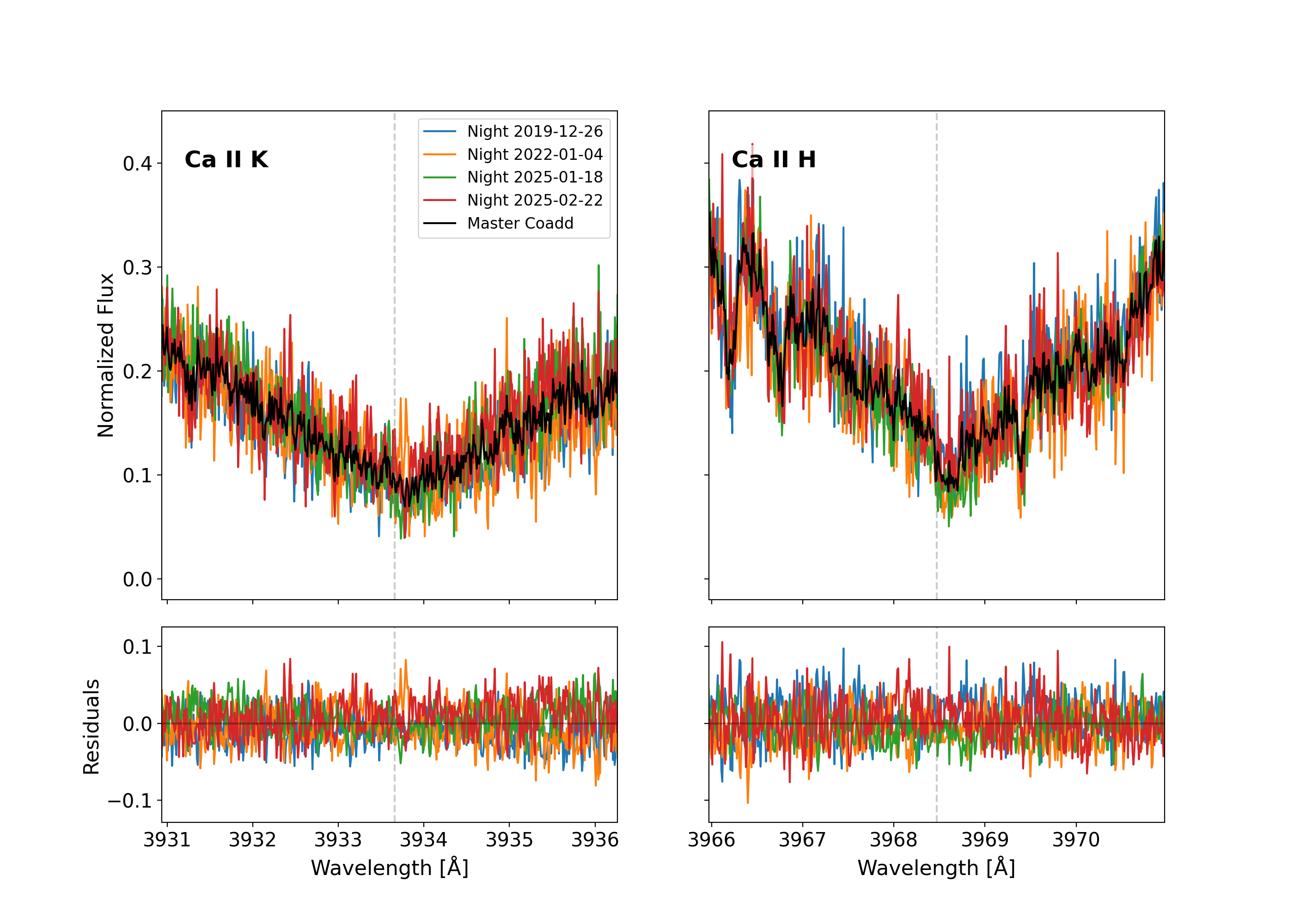}
    \caption{Co-added out-of-transit spectra of the \ion{Ca}{II} K (left) and H (right) lines for the four observing nights of WASP-13 (top panels) and their residuals relative to the master co-added spectrum (bottom panels). All spectra are normalised to the local pseudo-continuum and corrected for the barycentric Earth radial velocity.
    The vertical dashed lines indicate the rest-frame laboratory wavelengths of the line cores. The lack of core emission or coherent variability across all nights confirms the quiet chromospheric state of WASP-13.}
    \label{fig:caii_hk}
\end{figure}

\begin{figure}
    \centering
    \includegraphics[width=\linewidth]{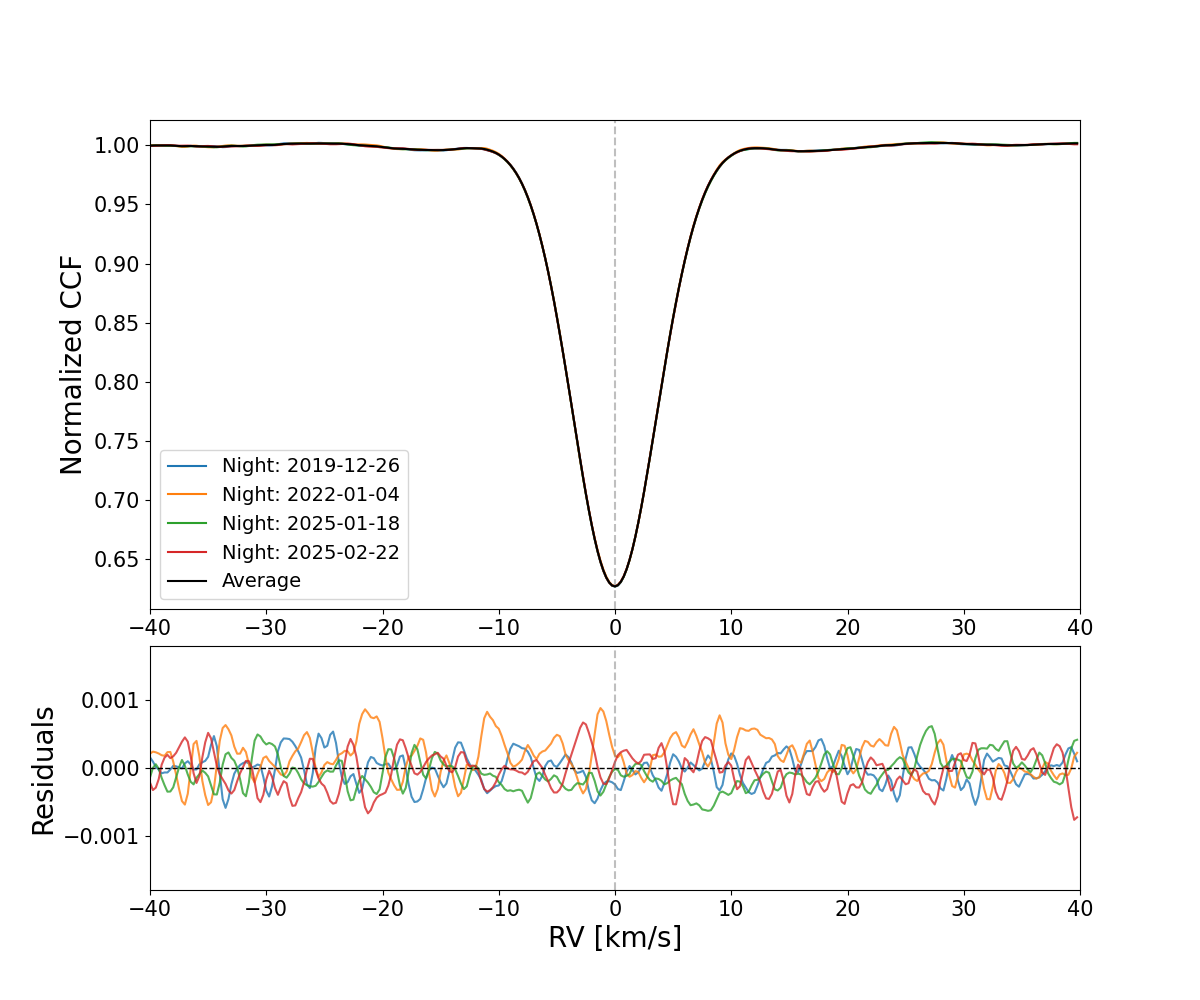}
    \caption{Out-of-transit \ac{CCF} linear profiles for WASP-13. \textit{Top}: Nightly average profiles extracted from the DRS using a G2 mask. The black line indicates the average out-of-transit \ac{CCF}. Profiles are centered in the stellar rest frame and normalised. \textit{Bottom}: Residuals relative to the average out-of-transit \ac{CCF}.}
    \label{fig:ccf_out}
\end{figure}

\begin{figure*}
\includegraphics[width=\linewidth]{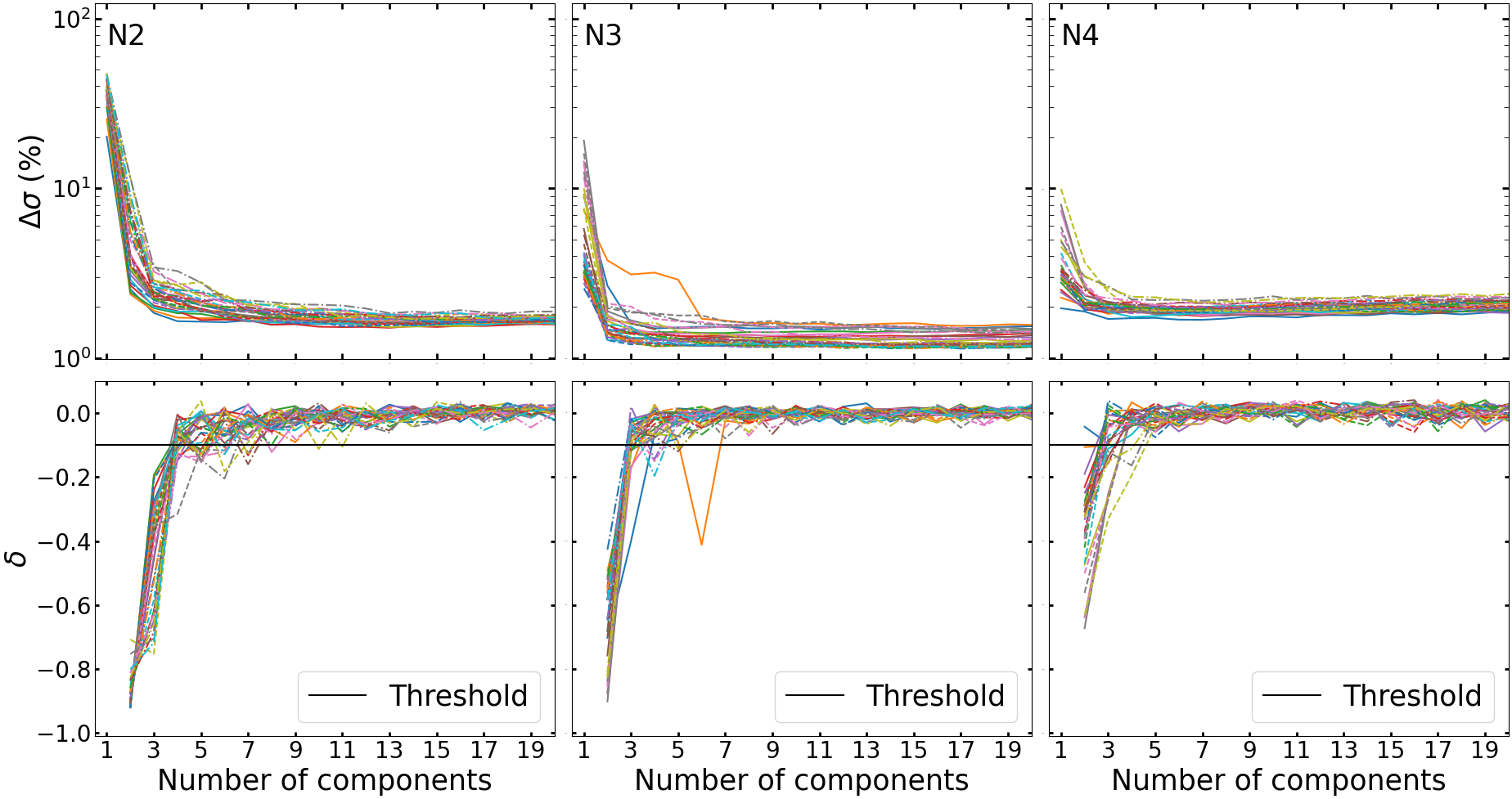}
\caption{Variation of the standard deviation ($\sigma$) of the residuals as a function of the number of principal components removed via PCA for the N2 dataset (left panels), the N3 dataset (central panels), and the N4 dataset (right panels). Top panels: relative variation of the standard deviation of the residuals ($\Delta\sigma=\frac{\sigma_{i-1}-\sigma_{i}}{\sigma_{i-1}}$, with $i$ = the $i$-th number of principal components removed), expressed in percentage value and reported in logarithmic scale for better visualisation, in the different spectral orders (coloured solid and dashed lines). Bottom panels: relative variation of $\Delta\sigma$ ($\delta$) in the different spectral orders (coloured solid and dashed lines). To find the minimum number of components at which the standard deviation of the residuals does not change in a significative way, we placed a threshold at $-0.1$ (solid black horizontal line), as in~\citet{basilicata2025}, and selected the minimum number of principal components for which $-0.1 < \delta < 0$ for all the spectral orders.}
\label{pca_components_plot}
\end{figure*}

\begin{table*}
    \centering
\caption{Time series of WASP-13\,b from HARPS-N data of night 2019 Dec 26 (N1). The BJD, \ac{RV}, and the $\log\,R^{\prime}_{\rm HK}$ values are shown with their related uncertainties.}\label{tab:rvs-rhk}
    \centering
    %\begin{spacing}{1.4}
    \begin{tabular}{c c c c c}
    \hline 
    \hline
        BJD & \ac{RV} [km$\,$s$^{-1}$] & $\sigma\rm_{RV}$ [km$\,$s$^{-1}$]& $\log\,R^{\prime}_{\rm HK}$ &$\sigma\rm_{\log\,R^{\prime}_{HK}}$\\
        \hline
        
2458844.50412695& 9.8815 & 0.0049& -5.3280 & 0.0785     \\
2458844.51482201& 9.8745  & 0.0045& -5.1834 & 0.0499 \\  
2458844.52566755& 9.8717  & 0.0047& -5.1787 & 0.0519     \\
2458844.53534404& 9.8716  & 0.0047& -5.0795 & 0.0412     \\
2458844.54682619& 9.8687  & 0.0051& -5.1065 & 0.0489     \\
2458844.55742866& 9.8796  & 0.0053& -5.1314 & 0.0535     \\
2458844.56799640& 9.8918  & 0.0052& -5.1293 & 0.0526     \\ 
2458844.57889981& 9.8877  & 0.0052& -5.0981 & 0.0484     \\
2458844.58962960& 9.8790  & 0.0047& -5.1378 & 0.0448     \\
2458844.60044041& 9.8755  & 0.0045& -5.1282 & 0.0417     \\
2458844.65698304& 9.8532  & 0.0039 & -5.1204 & 0.0336\\  
2458844.66730772& 9.8458  & 0.0035 & -5.1549 & 0.0308\\      
2458844.67966954& 9.8416  & 0.0038 & -5.1562 & 0.0352\\
2458844.69038776& 9.8458  & 0.0035 & -5.1946 & 0.0336\\
2458844.70122172& 9.8583  & 0.0044 & -5.1586 & 0.0421\\
2458844.71202095& 9.8576  & 0.0039 & -5.0991 & 0.0311\\
2458844.72249610& 9.8650  & 0.0044 & -5.1625 & 0.0439\\
2458844.73308699& 9.8512  & 0.0042 & -5.1325 & 0.0380\\
2458844.74341166& 9.8640  & 0.0055 & -5.1761 & 0.0622\\      
2458844.75516002& 9.8564  & 0.0041 & -5.1979 & 0.0436\\
 \hline
    \hline
    \end{tabular}
  %  \end{spacing}
\end{table*}

\begin{table*}
    \centering
\caption{Same as Table \ref{tab:rvs-rhk} but for the night of 2022 January 04 (N2).}\label{tab:rvs-rhk2}
    \centering
    %\begin{spacing}{1.4}
    \begin{tabular}{c c c c c}
    \hline 
    \hline
        BJD & \ac{RV} [km$\,$s$^{-1}$] & $\sigma\rm_{RV}$ [km$\,$s$^{-1}$]& $\log\,R^{\prime}_{\rm HK}$ &$\sigma\rm_{\log\,R^{\prime}_{HK}}$\\
        \hline
2459584.56944617& 9.8716 & 0.0053 & -5.0276     & 0.0524\\
2459584.57916882& 9.8825  & 0.0033 & -5.1120     & 0.0318\\
2459584.59058137& 9.8867  & 0.0036& -5.0843     & 0.0345\\
2459584.60164668& 9.8792  & 0.0028& -5.0802     & 0.0223\\
2459584.61124201& 9.8776  & 0.0031& -5.0873     & 0.0263\\     
2459584.62183276& 9.8738  & 0.0034& -5.1053     & 0.0329\\
2459584.63334947& 9.8650  & 0.0028& -5.1380     & 0.0247\\
2459584.64366243& 9.8604  & 0.0027& -5.0634     & 0.0197\\
2459584.65429948& 9.8545  & 0.0026& -5.0978     & 0.0206\\
2459584.66524904& 9.8504  & 0.0029& -5.1050     & 0.0268\\
2459584.67581664& 9.8388  & 0.0031& -5.1298     & 0.0304\\
2459584.68644211& 9.8370  & 0.0032& -5.1169     & 0.0316\\
2459584.69728750& 9.8322  & 0.0033& -5.1304     & 0.0347\\
2459584.70758888& 9.8396  & 0.0037& -5.1502     & 0.0425\\
2459584.71919819& 9.8594  & 0.0039& -5.0959     & 0.0412\\
2459584.72943013& 9.8535  & 0.0034& -5.0742     & 0.0322\\
2459584.74006717& 9.8472  & 0.0044& -5.0465     & 0.0430\\
2459584.75108618& 9.8571  & 0.0045& -5.1290     & 0.0548\\
2459584.76134126& 9.8521  & 0.0052& -5.0418     & 0.0546\\
2459584.77296214& 9.8512  & 0.0052& -5.1708     & 0.0727\\
 \hline
    \hline
    \end{tabular}
  %  \end{spacing}
\end{table*}

\begin{table*}
    \centering
\caption{Same as Table \ref{tab:rvs-rhk} but for the night of 2025 January 18 (N3).}\label{tab:rvs-rhk3}
    \centering
    %\begin{spacing}{1.4}
    \begin{tabular}{c c c c c}
    \hline 
    \hline
        BJD & \ac{RV} [km$\,$s$^{-1}$] & $\sigma\rm_{RV}$ [km$\,$s$^{-1}$]& $\log\,R^{\prime}_{\rm HK}$ &$\sigma\rm_{\log\,R^{\prime}_{HK}}$\\
        \hline
2460694.54566419& 9.8674  & 0.0033 & -5.0479     & 0.0259\\
2460694.55635887& 9.8713  & 0.0033 & -5.0948     & 0.0296\\
2460694.56734291& 9.8700  & 0.0028& -5.0973     & 0.0238\\     
2460694.57777138& 9.8785  & 0.0028& -5.0805     & 0.0225\\
2460694.58833874& 9.8820  & 0.0028& -5.1004     & 0.0234\\  
2460694.59926491& 9.8823  & 0.0029& -5.1473     & 0.0286\\
2460694.61013320& 9.8844  & 0.0029& -5.1097     & 0.0251\\     
2460694.62077001& 9.8836  & 0.0025& -5.0638     & 0.0182\\
2460694.63155729& 9.8778  & 0.0024& -5.0928     & 0.0179\\
2460694.64190474& 9.8763  & 0.0024& -5.0649     & 0.0159\\
2460694.65261099& 9.8695  & 0.0027& -5.0588     & 0.0185\\
2460694.66374549& 9.8636  & 0.0025& -5.0917     & 0.0179\\
2460694.67379201& 9.8498  & 0.0025& -5.0699     & 0.0179\\     
2460694.68432465& 9.8444  & 0.0030& -5.0761     & 0.0245\\
2460694.69579481& 9.8377  & 0.0032& -5.1316     & 0.0303\\
2460694.70638531& 9.8438  & 0.0030& -5.0800     & 0.0257\\ 
2460694.71706842& 9.8392  & 0.0029& -5.1230     & 0.0263\\  
2460694.72785569& 9.8463  & 0.0030& -5.0702     & 0.0242\\    
2460694.73837675& 9.8565  & 0.0030 & -5.0830     & 0.0250\\
2460694.74888624& 9.8512  & 0.0032 & -5.0585     & 0.0255\\     
2460694.75988185& 9.8518  & 0.0033 & -5.0570     & 0.0270\\
2460694.77053023& 9.8579  & 0.0034 & -5.0846     & 0.0303\\     
2460694.78142167& 9.8565  & 0.0030& -5.0836     & 0.0257\\
2460694.79188486& 9.8494  & 0.0027 & -5.0662     & 0.0204\\
\hline
    \hline
    \end{tabular}
  %  \end{spacing}
\end{table*}

\begin{table*}
    \centering
\caption{Same as Table \ref{tab:rvs-rhk} but for the night of 2025 February 22 (N4).}\label{tab:rvs-rhk4}
    \centering
    %\begin{spacing}{1.4}
    \begin{tabular}{c c c c c}
    \hline 
    \hline
        BJD & \ac{RV} [km$\,$s$^{-1}$] & $\sigma\rm_{RV}$ [km$\,$s$^{-1}$]& $\log\,R^{\prime}_{\rm HK}$ &$\sigma\rm_{\log\,R^{\prime}_{HK}}$\\
        \hline
2460729.36429265  & 9.8866 &  0.0033 & -5.0494   & 0.0278\\      
2460729.37458164& 9.8891 &  0.0028& -5.0577     & 0.0208\\
2460729.38531043& 9.8870 &  0.0027 & -5.1228     & 0.0226\\
2460729.39564572& 9.8836 &  0.0029& -5.0590     & 0.0219\\ 
2460729.40694161& 9.8896 &  0.0030& -5.0874     & 0.0249\\
2460729.41692969& 9.8959 &  0.0037 & -5.1108     & 0.0376\\      
2460729.42850335& 9.9074 &  0.0060& -5.0699     & 0.0686\\  
2460729.43809792& 9.8930 &  0.0049& -5.0916     & 0.0531\\
2460729.44905819& 9.8905 &  0.0040& -5.1053     & 0.0407\\      
2460729.45986799& 9.8979  & 0.0062 & -5.0522     & 0.0687\\
2460729.47141851& 9.8917  & 0.0038& -5.1706     & 0.0447\\
2460729.48208943& 9.8780  & 0.0036 & -5.0761     & 0.0328\\
2460729.49221638& 9.8692  & 0.0038 & -5.1383     & 0.0416\\  
2460729.50191512& 9.8726  & 0.0072& -5.1659     & 0.1081\\
2460729.51453041& 9.8568  & 0.0056 & -5.1131     & 0.0679\\
2460729.52395137& 9.8579  & 0.0078& -5.1209     & 0.1111\\      
2460729.53503894& 9.8523  & 0.0099& -5.3470     & 0.2540\\      
2460729.54567514& 9.8496  & 0.0104& -5.0578     & 0.1378\\
2460729.55714464& 9.8702  & 0.0076& -5.2500     & 0.1440\\   
2460729.56736419& 9.8697  & 0.0069& -5.1228     & 0.0936\\  
2460729.57830130& 9.8608  & 0.0069& -5.0591     & 0.0819\\
2460729.58810419& 9.8824  & 0.0058& -5.2143     & 0.0886\\
2460729.60031441& 9.8609  & 0.0044& -5.1536     & 0.0528\\      
2460729.61033720& 9.8675  & 0.0032 & -5.1063     & 0.0292\\       
    \hline
    \hline
    \end{tabular}
  %  \end{spacing}
\end{table*}

\begin{table*}
\centering
\begin{threeparttable}
\caption{GIANO-B spectral orders used in the analyses.}\label{tab:spectral_orders_gianob}
    \begin{tabular}{c | l | c}
    \hline
    \hline
\multicolumn{1}{c|}{Nights} & \multicolumn{1}{c|}{Discarded orders\tnote{\textit{a}}} &\multicolumn{1}{c}{CO restricted band\tnote{\textit{b}}}\\
      \hline
    N2 & [40-42, 54-56, 71-81] + [32, 62, 63] & [33] \\
    N3 & [40-42, 54-56, 71-81] + [46, 57, 62] & [32, 33] \\
    N4 & [40-42, 54-56, 71-81] + [46, 62, 68, 70] & [32, 33] \\
    N2+N3+N4 & [40-42, 54-56, 71-81] + [32, 46, 57, 62, 63, 68, 70] & [33]\\
    N3+N4 & [40-42, 54-56, 71-81] + [46, 57, 62, 68, 70] & [32, 33] \\
       \hline
    \end{tabular}
   \rule{0pt}{3ex}
\begin{tablenotes}\footnotesize
\item[\textit{a}] The orders that we discarded for the single-night and the multiple-night analyses. The latter are obtained by the union of the orders discarded in the single nights.
\item[\textit{b}] The orders used to search for CO, considering only the K band, which is covered by echelle orders 32 and 33 of GIANO-B. In this case, since in N2 the order 32 was discarded, only order 33 is considered in the analyses involving N2.
\end{tablenotes}
\end{threeparttable}
\end{table*}

\begin{figure*}
    \centering
    \includegraphics[trim=0 0 0 0,clip,width=0.35\linewidth]{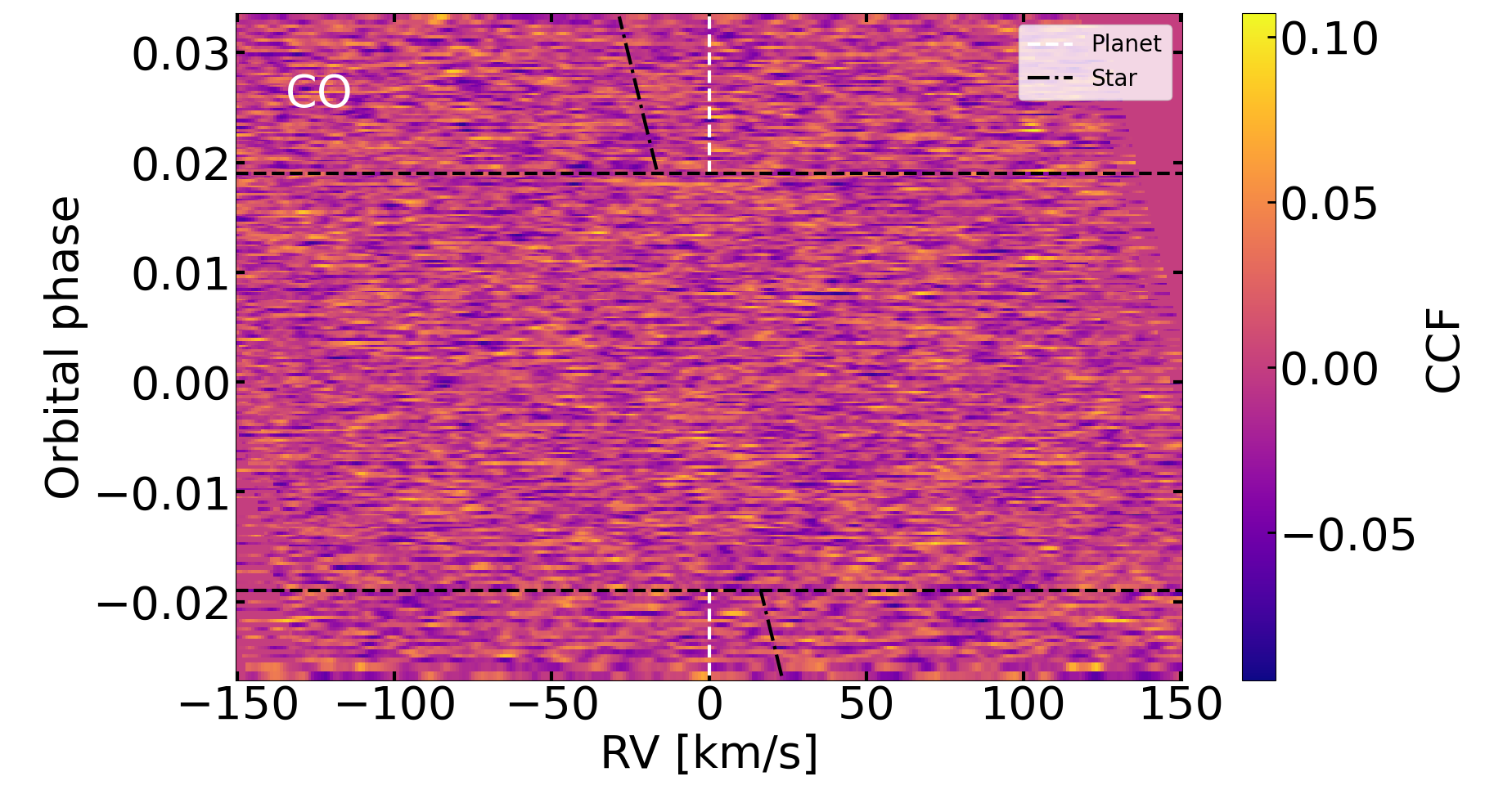}
    \includegraphics[trim=0 12 0 0,clip,width=0.3\linewidth]{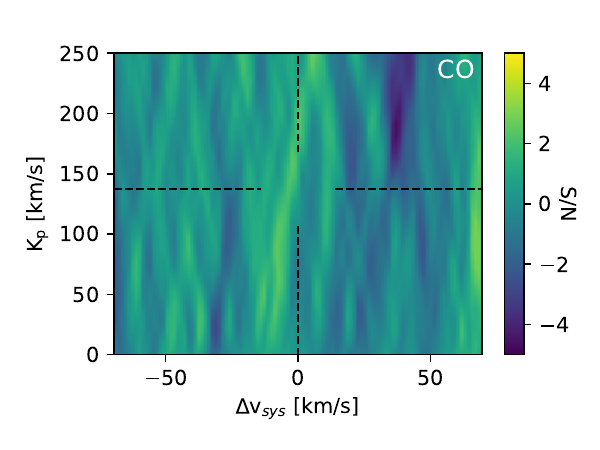}
    \includegraphics[trim=0 0 0 30,clip,width=0.3\linewidth]{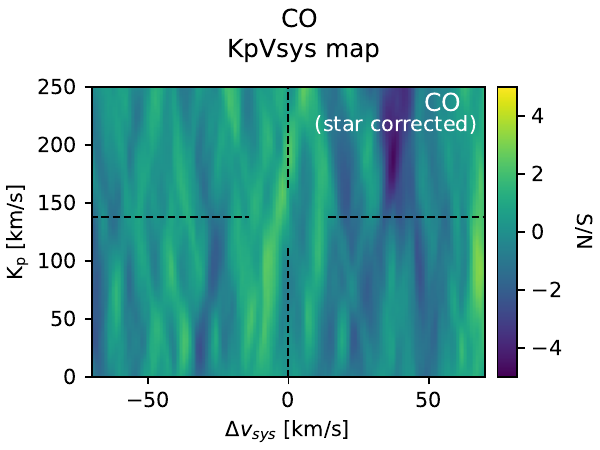}
    \caption{Cross-correlation maps obtained by combining the three NIR datasets for the CO template, using only the K band of GIANO-B. \textit{Left}: Cross-correlation as a function of the orbital phase in the planet rest-frame. The dashed white (dash-dotted black) line represents the planet (stellar) RV. \textit{Middle}: Plane of K$\rm_p$-$\Delta$\vsys \,, in which the dashed black cross indicates the expected K$\rm_p$ and $\Delta$\vsys \, of the system. \textit{Right}: Plane of K$\rm_p$-$\Delta$\vsys \, obtained after the subtraction of a realistic synthetic stellar spectrum to correct the RM and CLV effects in the K band of GIANO-B (see the text for more details).}
    \label{fig:map_Kp-vsys_co_order1}
\end{figure*}

\begin{figure*}
\includegraphics[width=\linewidth]{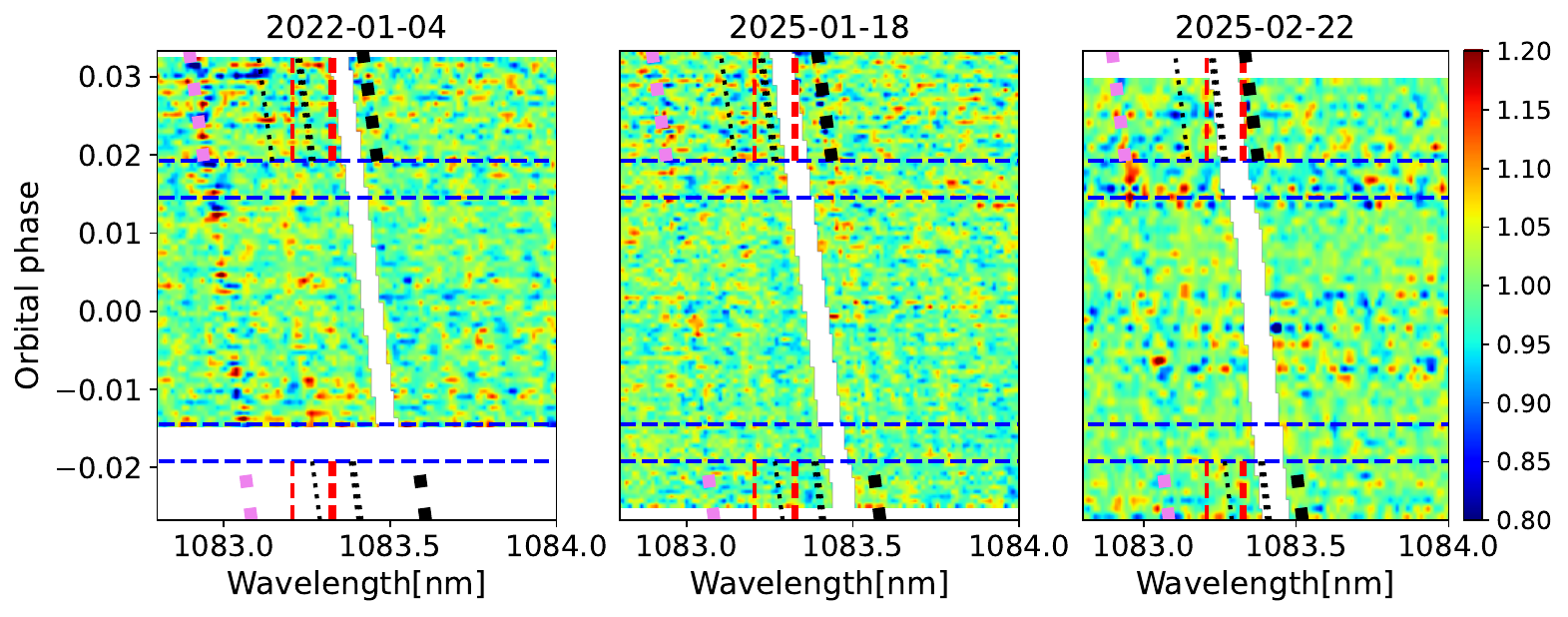}
\caption{Two-dimensional transmission spectra in the planet rest frame in the region of the \ion{He}{I} triplet, as a function of wavelength and planetary orbital phase. The contact points t$_1$, t$_2$, t$_3$ and t$_4$ are marked with horizontal blue lines. The regions affected by OH contamination are masked.  Some residuals are left at the position of the stellar Si line ($\sim$1083.0 nm, highlighted in pink). This is due to the depth of the line, which can cause difficulties in spectral extraction \citep[e.g. ][]{Krishnamurthy2023}. Dotted black and red lines mark the position of the \ion{He}{I} lines in the stellar and planet rest frame, respectively. Lines with black squares mark the position of the strong H$_2$O telluric line at $\sim$1083.5~nm. }
\label{mappe_Hei}
\end{figure*}

\begin{figure*}
    \centering
    \includegraphics[width=0.4\linewidth]{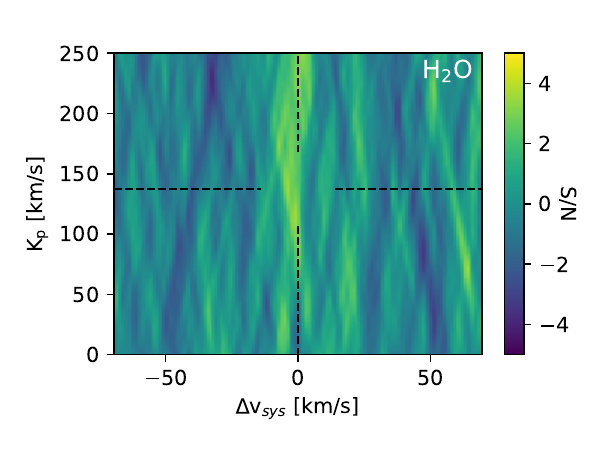}
    \includegraphics[width=0.4\linewidth]{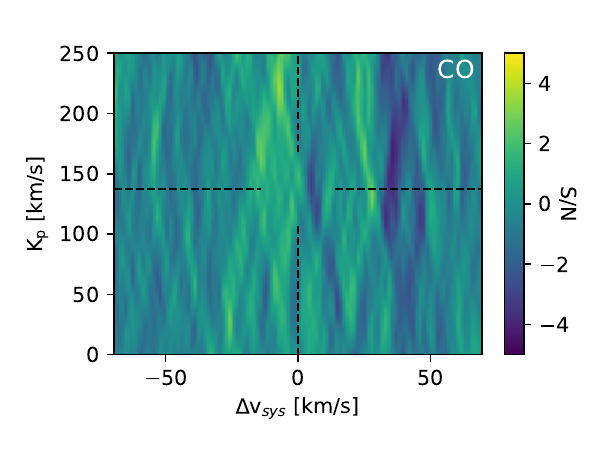}
    \caption{Planes of K$\rm_p$-$\Delta$\vsys \, obtained by combining all GIANO-B datasets for the H$_2$O and CO templates, after selecting a different number of components for the different spectral orders. %In each plot, the dashed black cross indicates the expected K$\rm_p$ and $\Delta$\vsys \, of the system.
    }
    \label{fig:3nights_free_tanlim}
\end{figure*}

\section*{Affiliations}
\noindent
{\it
$^{1}$INAF -- Osservatorio Astrofisico di Catania, Via S. Sofia 78, 95123 Catania, Italy\\
$^{2}$INAF -- Osservatorio Astronomico di Brera, Via E. Bianchi 46, 23807 Merate (LC), Italy\\
$^{3}$INAF -- Osservatorio Astrofisico di Torino, Via Osservatorio 20, 10025 Pino Torinese, Italy\\
$^{4}$INAF -- Osservatorio Astronomico di Palermo, Piazza del Parlamento, 1, 90134 Palermo, Italy\\
$^{5}$Department of Physics, University of Turin, Via Pietro Giuria 1, 10125 Torino, Italy\\
$^{6}$Université Côte d’Azur, Observatoire de la Côte d’Azur, CNRS, Lagrange, CS 34229, Nice, France\\
$^{7}$Space Research Institute, Austrian Academy of Sciences, Schmiedl- strasse 6, 8042 Graz, Austria\\
$^{8}$Department of Physics, University of Rome "Tor Vergata", Via della Ricerca Scientifica 1, 00133 Roma, Italy\\
$^{9}$Fundación Galileo Galilei - INAF, Rambla José Ana Fernandez Pérez 7, E-38712 Breña Baja, TF, Spain\\
$^{10}$INAF -- Osservatorio Astronomico di Roma, Via Frascati 33, I-00078 Monte Porzio Catone, Italy\\
$^{11}$University of Palermo, Department of Physics and Chemistry "Emilio Segrè", Via Archirafi 36, Palermo, Italy\\
$^{12}$Centro di Ateneo di Studi e Attività Spaziali “G. Colombo” -- Università degli Studi di Padova, Via Venezia 15, 35131 Padova, Italy\\
$^{13}$INAF -- Osservatorio Astronomico di Padova, vicolo dell’Osservatorio 5, I-35122 Padova, Italy
}

%\bsp	% typesetting comment
\label{lastpage} 
\end{document}